\documentclass[preprint]{aastex}

\slugcomment{To appear in Astronomical Journal}

\shorttitle{Radio Loud NLS1's}
\shortauthors{Whalen et al.}

\begin{document}

\title{Optical Properties of Radio-selected Narrow Line Seyfert 1 Galaxies}

\author{D. J. Whalen \altaffilmark{1,2}, 
S. A. Laurent-Muehleisen \altaffilmark{1},
E. C. Moran \altaffilmark{3},
R. H. Becker \altaffilmark{1,2}}

\altaffiltext{1}{Department of Physics, University 
of California, One Shields Avenue, Davis, CA 95616;
jwhalen@igpp.ucllnl.org}

\altaffiltext{2}{IGPP, L-413, Lawrence Livermore National Laboratory, 
Livermore, CA 94550; bob@igpp.ucllnl.org, slauren@igpp.ucllnl.org}

\altaffiltext{3}{Department of Astronomy, Wesleyan University, Middleton, CT, 
06459; ecm@astro.wesleyan.edu}

\begin{abstract}
We present results from the analysis of the optical spectra of 47
radio-selected narrow-line Seyfert 1 galaxies (NLS1s). These objects
are a subset of the First Bright Quasar Survey (FBQS) and were
initially detected at 20 cm (flux density limit $\sim1 $ \rm mJy) in
the VLA FIRST Survey. We run Spearman rank correlation tests on
several sets of parameters and conclude that, except for their radio
properties, radio-selected NLS1 galaxies do not exhibit significant
differences from traditional NLS1 galaxies. Our results are also in
agreement with previous studies suggesting that NLS1 galaxies have
small black hole masses that are accreting very close to the Eddington
rate. We have found 16 new radio-loud NLS1 galaxies, which increases
the number of known radio-loud NLS1 galaxies by a factor of $\sim 5$.

\end{abstract}


\section{Introduction}

Narrow line Seyfert 1 galaxies (hereafter NLS1s) are a class of Active
Galactic Nuclei (AGN) which, like Seyfert 1 galaxies, exhibit both
permitted and forbidden optical emission lines.  Unlike standard
broad-line Seyfert 1 galaxies, NLS1 galaxies additionally exhibit the
following observational properties \citep{goo89}: (1) the FWHM of the
H$\beta$ line is less than 2000 km s$^{-1}$, (2) permitted lines are
only slightly broader than the forbidden lines, (3) the forbidden-line
emission is relatively weak, i.e., [OIII]$\lambda5007$ / H$\beta <$ 3,
and (4) FeII and other high ionization emission line complexes are
unusually strong, i.e., stronger than ``normal'' Seyferts, especially
in the 4500$-$4680 \AA\/ and 5105$-$5395 \AA\/ wavelength ranges
\citep{ost85}.

It is not well understood at present what mechanism drives these
differences between normal broad-lined Seyfert 1 and NLS1
galaxies. Two opposing models exist, each with their own successes and
shortcomings. The first model suggests NLS1 galaxies are viewed with
the accretion disk in a face-on orientation. This case is similar to the Unified
 Model explanation for Seyfert 1 galaxies, with the crucial difference being that the emission lines appear narrower than typical Seyfert 1 galaxies because the optical line emitting
clouds are primarily confined to a disk perpendicular to the symmetry
axis, reducing the velocity dispersion along the line of sight
\citep{ost85,goo89, Puchnarewicz92,bol96}. The appearance of strong
FeII emission could also be explained by a pole-on geometry if the
FeII emission originates in the accretion disk as has been suggested
by \cite{fab89}.

It has been alternatively proposed that the unique properties
exhibited by NLS1 galaxies are a result of having accretion rates much
closer to the Eddington limit than ``normal'' broad-line Seyfert
galaxies \citep{lao97}.  Assuming that the efficiency of conversion of
accretion energy to radiation is the same in both near- and
sub-Eddington sources, those sources accreting near the Eddington rate
(NLS1s) should contain smaller mass black holes compared to
sub-Eddington sources of the same luminosity. Smaller mass black holes
in NLS1 galaxies would lead to narrower optical emission lines if the
motion of the BLR region clouds are dominated by Keplerian
velocities. This model also provides an explanation for many other
peculiar properties of the NLS1 galaxies. For example, when the
accretion rate approaches Eddington, the disk is likely ``puffed up''
by radiation pressure. This would result in a larger X-ray-heated
volume that can (a) generate the observed FeII emission and (b) shield
the NLR from UV ionizing radiation, resulting in weaker [OIII]
emission.

Additional fuel which favors the accretion rate model, is provided by
the work of \cite{bor92}. They measured many continuum and line
properties for a variety of AGN and examined various correlations and
inverse correlations between myriad parameters. They then reduced this
large set of data to a few principal components, eigenvectors of the
correlation matrix. The most important eigenvector is principal
component 1 (PC1) which links the strength of FeII emission, [OIII]
emission, and H$\beta$ line asymmetry, but is dominated by the inverse
correlation between the strengths of FeII and [OIII]. The most popular
interpretation for what PC1 physically represents is a sequence in
L/L$_{\rm Edd}$, the Eddington ratio. The idea is that the vertical
structure of the accretion disk, governed by the Eddington ratio,
drives line strengths and continuum components through its
illumination of broad-line clouds and an extended narrow-line
region. NLS1 galaxies lie at one extreme of the PC1 correlation
(strong FeII / weak [OIII]), lending support to the idea that near
Eddington accretion rates (around small mass black holes) ultimately
determine whether an object manifests itself as a NLS1 or Seyfert 1
galaxy.

But more than optical emission line properties distinguish NLS1
galaxies from other classes of AGN; their X-ray properties also show
marked differences compared to other types of Seyfert galaxies. The
best known X-ray property of NLS1 galaxies is the presence of a soft
X-ray excess \citep{bra94}). NLS1 galaxies also exhibit unusually
strong X-ray variability \citep{bol97,lei99}.  Finally, NLS1 galaxies
are known to generally exhibit steeper hard (2-10 keV) X-ray spectra
than broad-line Seyfert 1 galaxies \citep{BMA97}.

Most of these X-ray properties can be explained within the framework
of either the orientation or accretion-rate models, although each
requires some special assumptions or modifications.  Within the
general confines of the orientation model, \cite{madau88} has examined
the properties of geometrically thick accretion disks and predicts a
strong soft X-ray emission when these disks are viewed face-on. In the
accretion-rate models, the observed soft excesses in NLS1 galaxies is
attributed to thermal emission from a viscously heated accretion disk
that results at an accretion rate close to the Eddington limit
\citep{pou95}. Likewise, enhanced variability in NLS1 galaxies can be
explained via either the orientation or accretion rate models. If NLS1
galaxies do accrete near Eddington and have smaller mass black holes,
their size scale is smaller and the variability can be more rapid. In
the orientation model, the variability may be enhanced via
relativistic boosting of the emission.

The trend toward harder spectra in normal broad-lined Seyfert galaxies
is the one observational characteristic which is difficult to explain
in the orientation model. In this model, the hard photon index is
expected to become softer as the inclination angle increases
\citep{haardt93}, the opposite of which must be true for this model to
explain differences seen between NLS1 and normal Seyfert 1
galaxies. However, when this model is generalized, and the hot phase
is assumed to be in localized regions on the disk, the dependence on
inclination decreases, and the index depends rather on the height of
the hot phase blobs above the disk. In this case, there is a need to
invoke a mechanism which links the height of sites which produce the
bulk of the observed emission with orientation in order to fit within
the orientation based scheme. This is a difficult proposition without
invoking some ad hoc physical structure that reflects or absorbs
radiation asymmetrically. In the high accretion rate model, however,
the steeper hard X-ray spectrum of NLS1 galaxies can by explained if
the hot phase luminosity is smaller in NLS1 galaxies, presumably due
to the smaller central black hole mass.

Despite years of work, both the orientation and accretion rate models
are still essentially equally viable; both provide a framework which
unifies NLS1 galaxies and normal Seyfert 1 galaxies, yet neither is entirely
satisfactory.  Additional observational constraints would go a long
way toward resolving the question. Radio observations may well provide
those constraints; they have undoubtedly been crucial in the
formulation of unification scenarios for other classes of AGN. But
while Seyferts of all types have been studied extensively at optical
and X-ray wavebands, little is known about the radio properties of
NLS1 galaxies in part because they are typically faint radio sources, if they
are known to be radio sources at all. One of the few systematic
studies of the radio properties of NLS1 galaxies is by
\cite{ulv95}. They observed a group of 15 NLS1 galaxies (of which only 9 were
detected in radio) and found relatively modest radio powers for their
sample (10$^{27}$ - 10$^{30}$ erg s$^{-1}$ Hz$^{-1}$). These 15 NLS1 galaxies
were all radio quiet, i.e., the ratio of radio luminosity to optical
luminosity is relatively small. \cite{mor03} show in their study that NLS1
galaxies can be radio intermediate and even radio luminous, and, as a
whole, are more radio luminous than classical Seyferts.

The work of \cite{ulv95} is a radio study of optically selected NLS1
galaxies while the \cite{mor03} work is a radio study of X-ray and IR
selected NLS1 galaxies.  There are no studies of a radio-selected
sample of NLS1 galaxies. This is partly because only recently has it
become apparent that NLS1 galaxies can exhibit significant radio
luminosities \citep{siebert99,gliozzi01,osh01}. Additionally, it
is only recently that new radio surveys have emerged which cover a
wide area of sky to depths of only a few mJy. It is from one such
survey, the FIRST (Faint Images of the Sky at Twenty Centimeters;
\cite{bec95}) that we have compiled the first known sample of
radio-selected NLS1 galaxies, and in the process increased by a factor
of five the number of known radio-loud NLS1 galaxies.

In this paper we present 47 NLS1 galaxies along with 15 BLS1 galaxies
with very narrow lines selected from FIRST and analyze their optical
and radio properties. We will use an H$_{0} = 65$ km s$^{-1}$,
$\Omega_{matter} = 0.3$, $\Omega_{\Lambda} = 0.7$ cosmology
throughout.

The structure of this paper is as follows: In sections 2 and 3 we
describe the sample selection criteria. In section 4 and 5 we describe
the data and the data analysis and in section 6 the method we use for
estimating central black hole mass in explained. We present our
results and correlations in Section 7 and discuss the radio properties
in section 8. The conclusion follows in section 9.
\section{The Sample}

Our objects were initially selected from the FIRST radio survey which
at the time (Feb.  1999) covered $\sim$ 2700 square degrees of the sky
centered on the North Galactic Cap, and is sensitive down to a flux
density limit of $\sim$ 1 mJy at a wavelength of 20 cm. The FIRST
survey was carried out with the VLA in the B-configuration and has a 5
arcsecond resolution.

The present sample of NLS1 galaxies is a subset of the larger FIRST Bright 
Quasar Survey (FBQS) and therefore shares many of the FBQS selection 
criteria, the details of which can be found in \cite{whi00}. Briefly, the 
digitized Palomar Observatory Sky Survey (POSS) was searched for optical 
counterparts within 1.2 arcseconds of FIRST radio sources. If the optical 
counterpart existed and was classified as stellar by the Automated Plate 
Machine (APM) on either the red (E) or blue (O) plates, the object became an 
FBQS candidate. Objects brighter than 17.8 mag on the E-plate (using 
re-calibrated, extinction corrected magnitudes), with colors bluer than 
(O - E) $=$ 2 were selected for spectroscopic followup observations. Optical 
spectroscopy was then performed on over 1,200 candidates.

Among the many quasars, low luminosity AGN, starburst and ``normal''
galaxies in the original FBQS study, 45 objects were selected as
possible NLS1 galaxy candidates. An additional 17 objects were
obtained from various followup programs to the FBQS, including the
partially complete (at the time) ``zero declination'' strip, a small
multiply-observed (in the radio) area of sky centered at zero degrees
declination and extending from approximately 21 hours to 3 hours of
right ascension. Only the initial 45 objects are from a spatially
complete sample, having been selected as part of the original 2,700
square degree survey. All objects do, however, satisfy the above
criteria. 

\begin{figure}
\plotone{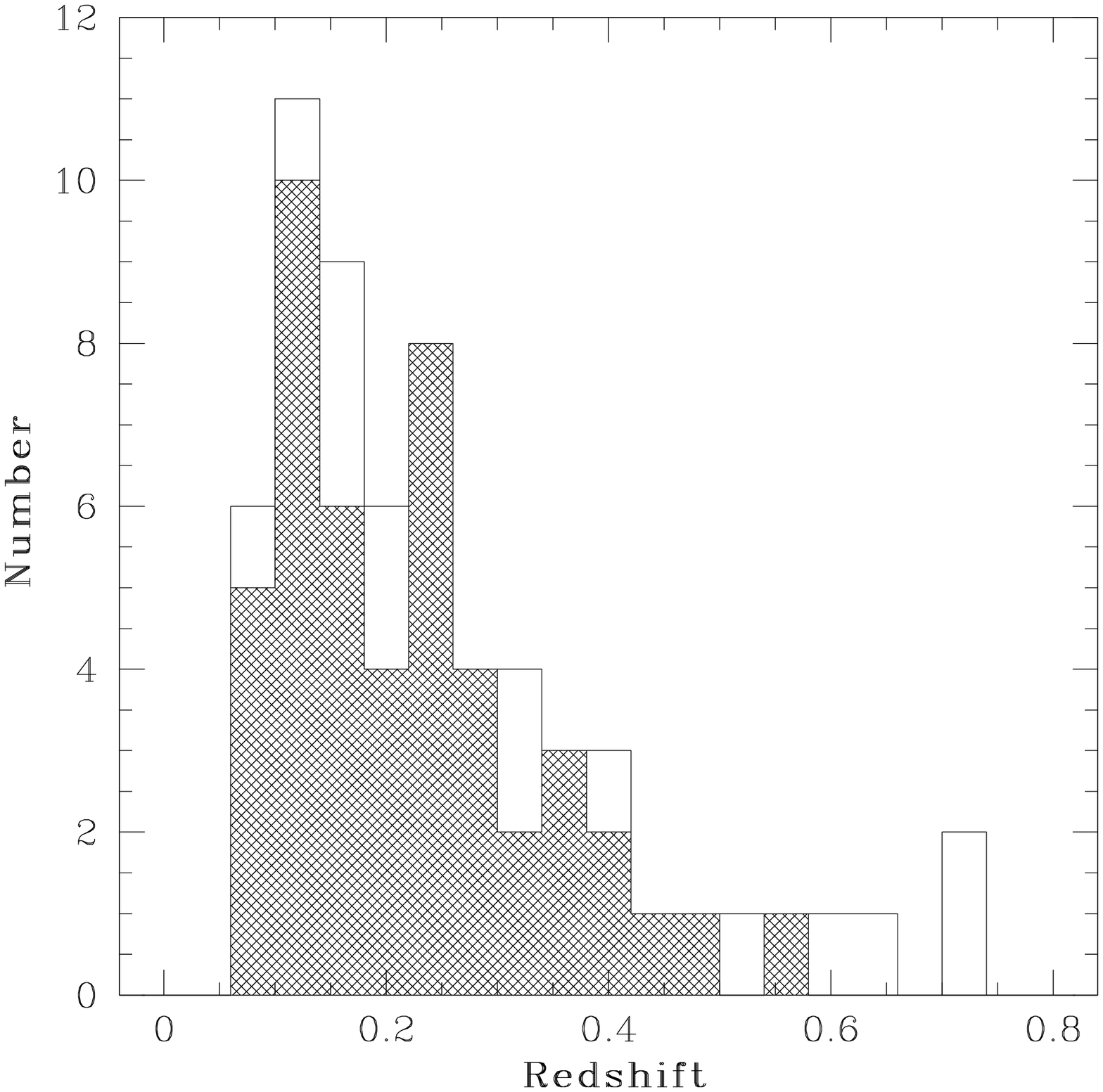}
\caption{The distribution of redshifts for our sample of FBQS
NLS1s. The shaded region represents the NLS1 galaxies while the other
objects are BLS1 galaxies. Note the dearth of nearby NLS1s. This is
due to FBQS selection criteria. If an object's optical image was
resolved, it was removed from the FBQS.
\label{fig1}}
\end{figure}

The sample of 62 NLS1 and BLS1 galaxies has a redshift range extending
from z = 0.065 to 0.715.  The distribution of redshifts for all 62
galaxies is shown in Figure \ref{fig1}. Due to the fact that these
objects were primarily identified by H$\rm \beta$ FWHM and [OIII]
properties, one would not expect to find objects of much greater
redshift in this sample, as the H$\beta$ and [OIII] lines would be
redshifted out of the optical wavelength range.



We note that one NLS1 galaxy in our sample, 0833$+$5124, actually
violates one of the FBQS selection criterion, having an extinction
corrected E-magnitude of 17.85. We nevertheless include it here, since
we intend to find NLS1 galaxies with radio emission. We additionally
point out that many of our spectra have FWHM($H\beta$) $> 2000$ km
s$^{-1}$. In the original definition for NLS1 galaxies\citep{ost85},
the criteria for line width was simply that the permitted lines were
only slightly broader than forbidden lines. We have decided to include
these galaxies as we can investigate these transition objects along
with our NLS1 galaxies. We shall refer to these objects as Broad Line Seyfert 1 galaxies (BLS1s) consistently. None of these objects have very broad lines,
the largest being $\sim 3000$ km s$^{-1}$ full width at half
maximum. However, these transition objects shall always be clearly
marked in the figures and any correlations will be run both with and
without these transition objects.

We also note that this sample should {\em not\/} be viewed as a
complete sample of radio-selected NLS1 galaxies, since the inclusion in the
FBQS requires the optical counterparts be unresolved and not very red;
clearly these criteria select against Seyfert type galaxies.
Nevertheless, significant numbers of Type 1, 2, Narrow-line Seyfert
and starburst galaxies are present in the FBQS. This is because (1)
the morphology classification from the APM plates are not precise and
(2) the FBQS selection criteria erred on the side of caution,
excluding only {\em obviously\/} resolved and/or red objects. Still,
our sample of NLS1 galaxies is incomplete with respect to a purely
radio-selected NLS1 sample, but will provide a useful comparison
between the class properties of optical- vs.  radio-selected NLS1
galaxies.

This comparison is particularly enlightening since PC1 tends to also
trace degrees of radio-loudnesses (ratio of radio to optical
flux). Objects that have strong FeII and weak [OIII] also tend to be
radio-quiet and to have H$\beta$ lines that are narrower; objects that
have weak FeII and strong [OIII] tend to be radio-loud and to have
H$\beta$ lines that are broader. Narrow-line Seyfert 1 galaxies,
therefore are expected to be radio-quiet (and usually are). However,
by selecting our sample from FIRST, we favor finding radio-loud
objects which most strain the PC1 correlation. In fact, many of the 47
narrow line Seyfert 1 galaxies in our sample are radio-loud, some
remarkably so (See section 5). While it is not surprising we have
found some radio-loud NLS1 galaxies, it is interesting that we have found
enough to increase by a factor of five the number of known radio-loud
NLS1 galaxies. These objects in particular should test the competing merits of
the orientation vs. accretion-rate models.

\section{NLS1/BLS1 Cut}

\begin{figure}
\plotone{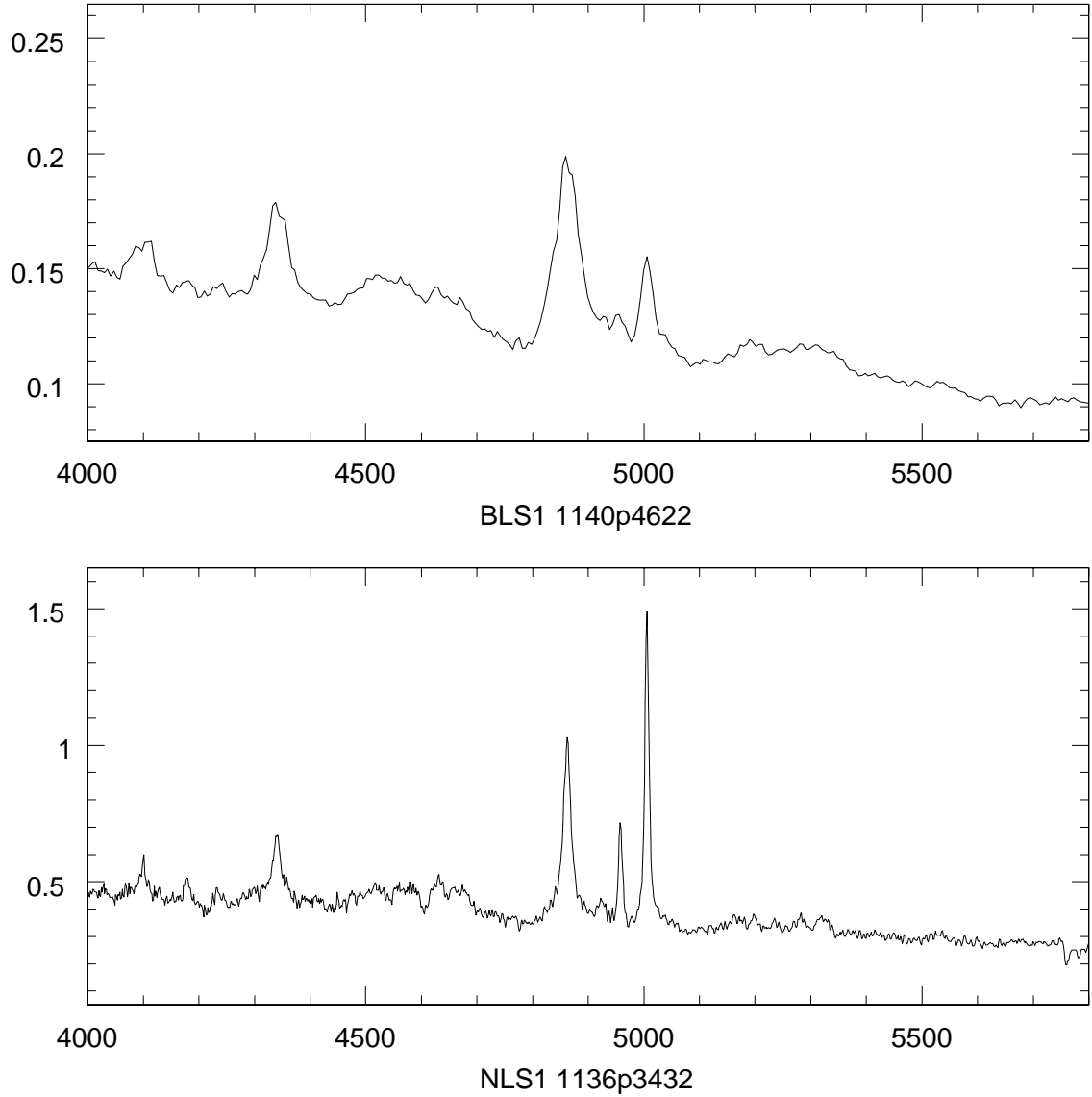}
\caption{An example of an NLS1 and a BLS1. In the upper panel is 1140p4622 whose FWHM($H\beta$) $=  2332$ km s$^{-1}$ which qualifies as a standard Seyfert 1 galaxy (although on the narrow side). In the lower panel is 1136p3432 which has FWHM($H\beta$) $= 918$ km s$^{-1}$. The horizontal axis has units of angstroms and the vertical axis has units of $10^{-14}$ $ \rm erg$ $\rm  s^{-1}cm^{-2}\AA^{-1}$.
\label{fig2}}
\end{figure}

\begin{figure}
\plotone{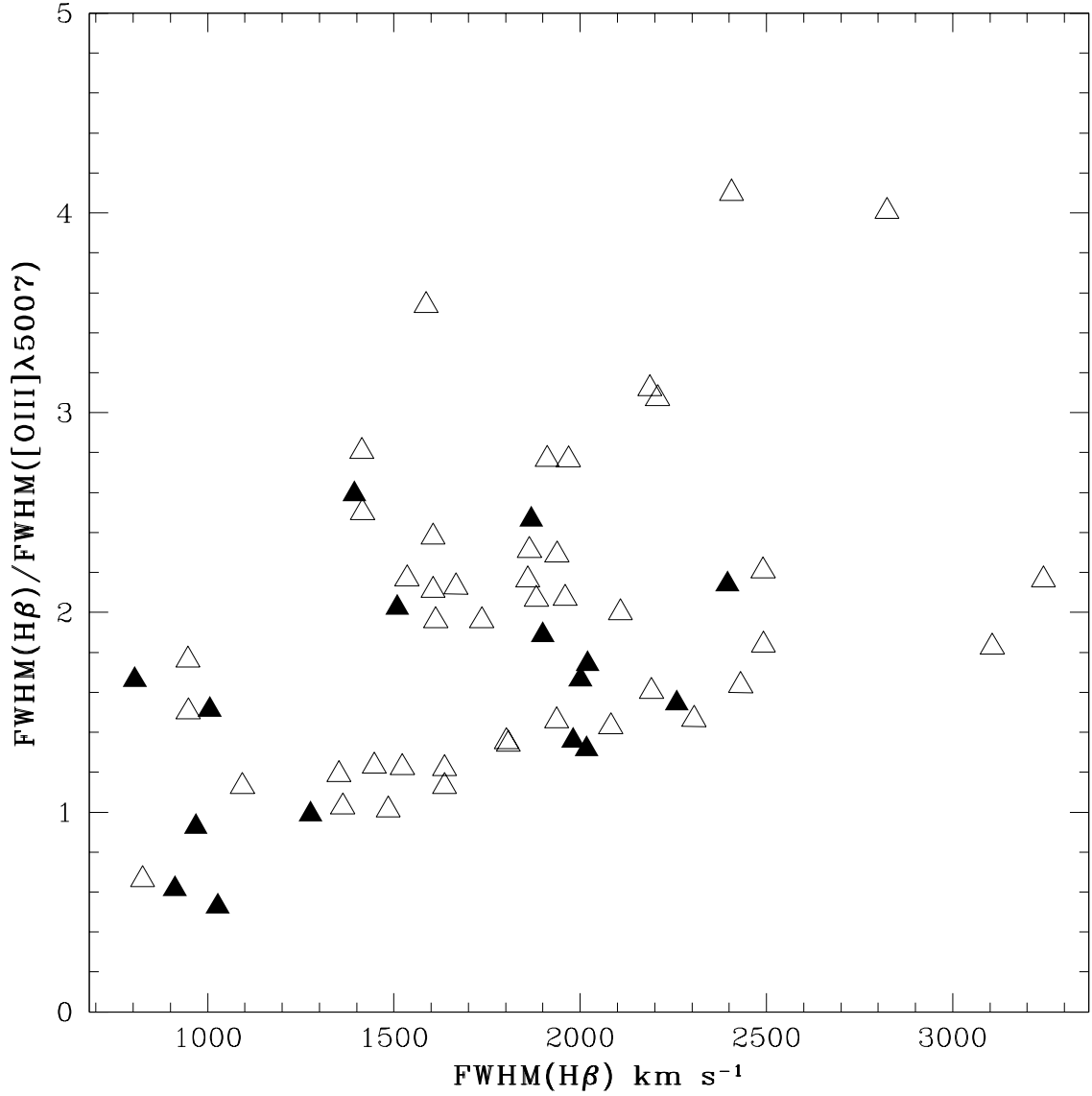}
\caption{The ratio of FWHM(H$\beta$)/FWHM([OIII]) plotted against
FWHM(H$\beta$). Shaded triangles represent the radio loud objects, open triangles represent the radio quiet objects. As seen
in the figure, the ratio for the objects with FWHM(H$\beta$)$>\sim
2000$ km s$^{-1}$ are not statistically different from the traditional
NLS1s with FWHM(H$\beta$)$<\sim 2000$ km s$^{-1}$.
\label{fig3}}
\end{figure}

In this section we address which objects should be considered NLS1
 galaxies, and which should be standard Seyfert 1
 galaxies, which we will refer to as BLS1s as mentioned previously. An example of each is shown in Figure \ref{fig2}. How should we make the cut? The criteria of \cite{goo89} state
 that FWHM (H$\beta$) should be less than $2000$ km s$^{-1}$, but
 Osterbrock \& Pogge's 1985 criteria were simply that FWHM(H$\beta$)
 only slightly greater than the width of the forbidden lines. As the
 attentive reader may have noticed, some of our objects have a
 FWHM(H$\beta) > 2000$ km s$^{-1}$. Why do we include these? These
 objects have FWHM(H$\beta)$ whose widths are not much greater than
 the FWHM([OIII]) as we can be seen in Figure \ref{fig3}. In this
 figure we plot the ratio FWHM(H$\beta)$/FWHM([OIII]) against
 FWHM(H$\beta)$. When we compare the objects less than $2000$ km
 s$^{-1}$ and greater than $2000$ km s$^{-1}$, there is little
 difference. Other than the two objects whose ratio is $\sim 4$ at
 about $2500$ km s$^{-1}$, the objects whose FWHM(H$\beta$) are
 greater than $2000$ km s$^{-1}$ are not unusual. Considering that
 these objects satisfy NLS1 criteria in all other ways,
 ([OIII]/H$\beta$ $< 3$, significant FeII complexes, small black hole
 mass), it doesn't seem unreasonable to call these objects NLS1
 galaxies. We shall not be taking such a bold step, however, and will
 simply refer to objects with H$\beta > 2000$ km s$^{-1}$ as Broad Line
 Seyfert 1 galaxies, or BLS1s. Objects with H$\beta < 2000$ km s$^{-1}$
 will be labelled NLS1s, and are the traditional narrow-line Seyfert 1
 galaxies. X-ray data on these objects could define where the physical
 break between NLS1s and BLS1s should be, if it exists.

\section{The Data}

Once FBQS targets were identified, spectra of the candidate quasars
were taken at several different sites and of the initial $\sim$ 1200
point sources in the FBQS \citep{whi00}, $\sim$ 600 were classified as
starburst galaxies with the rest being classified as Active Galactic
Nuclei (AGN). The NLS1 candidates were chosen through visual
inspection and then confirmed using the above criteria, i.e., H$\beta$
FWHM similar to forbidden lines, FeII emission was present, and
[OIII]/H\rm{$\beta$} $<$ 3. The spectra of all 62 objects can be found in the FBQS papers, \cite{gre96} and \cite{whi00}.

Six observatories were used to obtain the spectra. The telescopes and
instruments and the respective instrumental resolutions appear in Table \ref{tbl-1}. 
 The observations were made under a wide
variety of conditions, from photometric to cloudy, with both good and bad seeing. Many of the
objects were observed twice. The spectral data were reduced using
standard long slit spectral reduction packages in IRAF.

\section{Analysis}
 
The reduced data were fit using the SPECFIT routine in the IRAF
(v.2.11.3) data analysis program. The broad permitted lines of the
FeII complexes commonly found in NLS1 galaxies can severely contaminate the
spectrum and make accurate measurements of fluxes and line widths
difficult. In an effort to increase the accuracy of the line profiles,
the FeII complexes were subtracted from the data before measurement
of the other quantities. We used the method of \cite{bor92} to remove
the FeII complexes, a brief description of which follows. 

The FeII complexes from the spectrum of I Zw 1 were used to create a
template for use in modeling the FeII emission of other NLS1, the same
procedure and template used by Boroson \& Green(1992). The FeII template has
spectral coverage extending from $\lambda$4250 $ $\AA $ $ to
$\lambda$5700$ $ \AA. We varied the amplitude of the template and
also Doppler broadened the line profile until we were able to fit the
observed FeII lines. We then subtracted the FeII contribution to our
spectra. All data were de-redshifted to the rest frame before being
fit.

\begin{figure}
\plotone{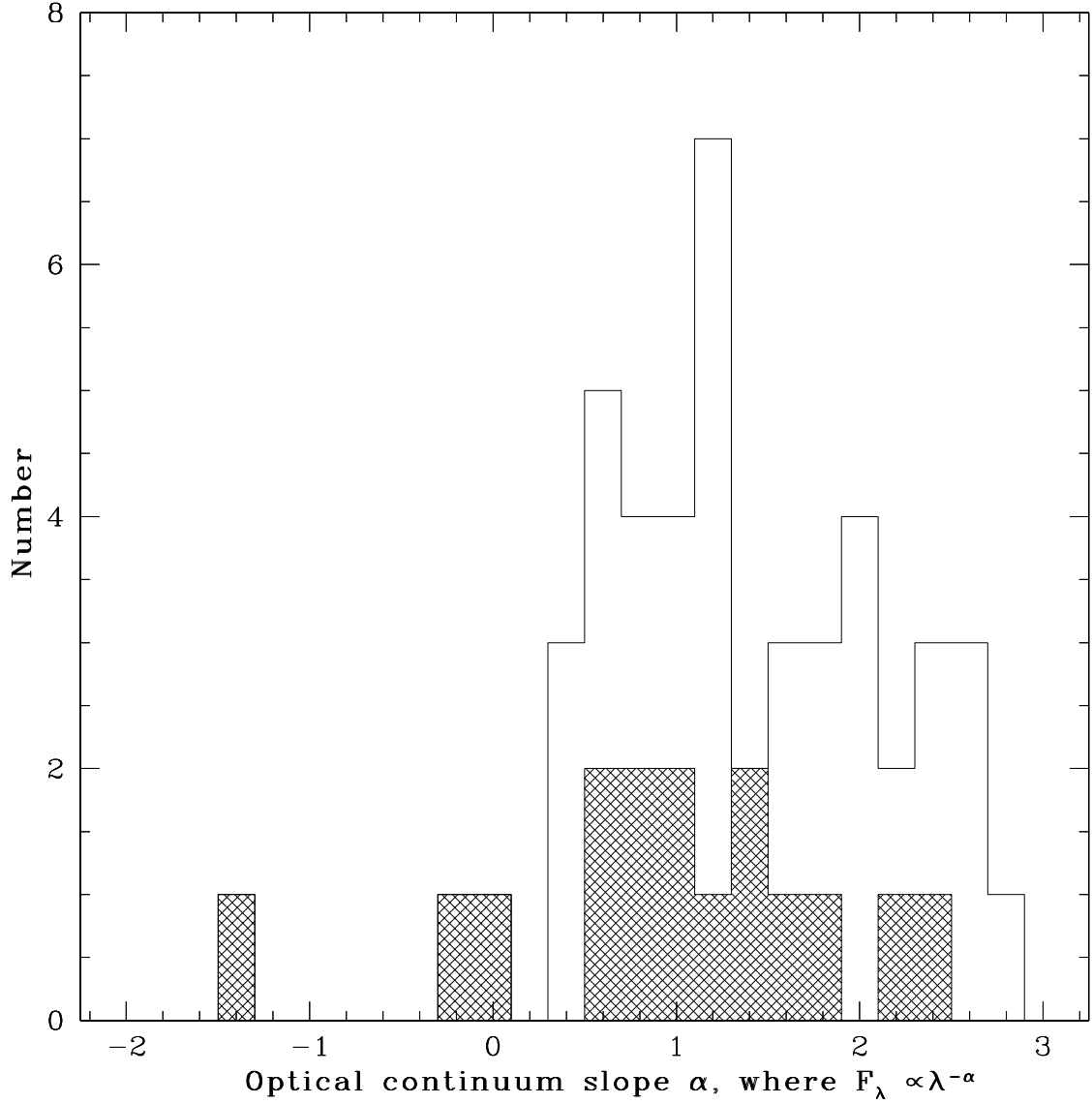}
\caption{The distribution of the optical continuum slopes, $\alpha$,
for the continuum from $3000$ \AA$ $ to $7000$ \AA $ $ for the NLS1
galaxies only. The shaded region represents the radio loud
distribution, the open region the radio quiet distribution. Spectra were fit with a powerlaw of the form
$F_{\lambda} \propto \lambda^{-\alpha}$.
\label{fig4}}
\end{figure}

The  H$\alpha$, H$\beta$, and H$\gamma$ emission lines were fit both
with a Lorentzian profile and a Gaussian profile. We observed that the Gaussian
profile missed the broad wings of the hydrogen lines, confirming
previous results (i.e.,\cite{ver01} and references therein). The flux,
FWHM and equivalent width of each line was measured. The
[OIII]$\lambda$5007 and [OIII]$\lambda$4959 forbidden lines were fit
using a Gaussian line profile. The FWHM for the emission lines
H$\beta$, [OIII]$\lambda$5007, and [OIII]$\lambda$4959 for each source
were measured. All line profiles have been corrected for instrumental broadening, using the instrumental resolution calculated by \cite{whi00}, for which the authors used calibration lamp lines to determine the broadening.
These quantities are tabulated in Table \ref{tbl-2}
with the exception of [OIII]$\lambda$4959 whose FWHM and flux were
forced to follow [OIII]$\lambda$5007, i.e., the FWHM is the same and
the flux is one third the [OIII]$\lambda$5007 value. Three objects,
1421+2824, 1644+2619, and 2155$-$0922, have sky absorption lines that
eliminated significant parts of the [OIII]$\lambda$5007 line. We were unable to accurate measurements of these three lines due to their destruction by the sky lines. [OIII]$\lambda4959$ was either too small or also affected by the sky lines to use as an estimate. The three cases are
noted in Table \ref{tbl-2}. The continuum portion of the spectrum was fit with a
power law (F$_{\lambda}$ $\propto$ $\lambda ^{-\alpha}$). The optical
continuum slopes are defined between 3000 \AA$ $ and 7000 \AA$ $ and
the distribution of optical continuum slopes, $\alpha$, is shown in
Figure \ref{fig4}.


The observing conditions were not necessarily photometric and
therefore we restrict the bulk of our analysis mainly to quantities that
are not dependent on absolute flux scale. Equivalent widths for the
emission lines were calculated using the measured integrated flux from
SPECFIT and dividing by the continuum strength at the center of the
line using the power law model fit that SPECFIT produced. To get an
equivalent width for the FeII continuum, we restricted our attention
to two portions of each individual spectrum. The section between 4434
\AA $ $ and 4684 $ $ \AA $ $ and from $5147$ $ $ \AA$ $ to 5350 $ $
\AA $ $ was used except in the case of 1256+3852. For this object we
considered the blue section only as the wavelength coverage was not
great enough to reach the red section. The H$\beta$,
[OIII]$\lambda$5007, [OIII]$\lambda$4959, and H$\gamma$ emission lines
were subtracted and the remaining spectra were normalized by the
continuum. We then integrated the FeII spectrum over the two sections
to get the equivalent width of the entire FeII complex (see Table
\ref{tbl-2}). 

The k-corrected radio loudness parameter, R$^*$, was
calculated for each object using the formulas \citep{sto92},

\begin{equation}
\log \rm f(5\, GHz)  =  -\, 29.0 + \log{ S_{\nu}} 
   +\ \alpha_{\nu} \log{[5/\nu]} 
-\, (1 + \alpha_{\nu}) \log{(1 + z)}          , 
\end{equation}

\begin{equation}
\log \rm f(2500 \AA) = -22.38 - 0.4\rm{B} 
   +\, (1 -\alpha_{\lambda})\log(1 + z) 
  +\, ( 2 - \alpha_{\lambda})\log(\frac{2500 \AA}{4409 \AA}),
\end{equation}

\begin{equation}
\log \rm R^{*} = \log f(5\,  GHz) - \log f(2500  \AA).
\end{equation}

In Equation 1 above, f(5GHz) refers to the
5GHz (6cm) flux, S$_{\nu}$ is the flux of frequency $\nu$, and $\nu$ is
the frequency in units of GHz.  The symbol $\alpha_{\nu}$ is the slope of the radio
continuum such that S$_{\nu} \propto \nu^{\alpha}$ and z is the object's redshift. In Equation 2 $\rm f(2500 \AA) $ is the rest frame flux at $2500 \rm \AA$, z is the redshift, B is the Johnson B-band magnitude, and
$\alpha_{\lambda}$ is the slope of the optical continuum where
$F_{\lambda} \propto \lambda^{-\alpha}$. 
Formula 2 has been altered from the version in \cite{sto92}. That
formula had no optical slope variable, $\alpha_{\lambda}$. Instead, a slope of $\alpha_{\lambda} =
-1$ (in our form of the equation) was assumed and hard-wired into the formula. We have measured
optical slope as part of our analysis so we followed \cite{sch68},
\cite{wil78}, and \cite{sra80}, and put the optical slope variable
back into the equation. We follow the procedure laid out in
\cite{whi00} and substitute O magnitudes from POSS I for Johnson
B-band magnitudes.

A histogram of the optical slope data, gathered in bins of size $0.20$
is shown in Figure \ref{fig4}, with the radio loud objects being the
shaded part of the distribution. We use Formula 2
instead of simply getting the $2500$\AA $ $ flux from our spectra for two reasons. The first reason is that all of our objects have O magnitudes while not all of the spectra have the coverage to reach $2500$ \AA. The second is
that the POSS magnitudes have been recalibrated plate by plate using
magnitudes from the Minnesota Automated Plate Scanner POSS-I catalog
and carefully corrected for extinction by the FBQS team (\cite{whi00}
and references therein), while our spectra are not assumed to be
photometric and therefore not as useful for calculation requiring
absolute flux scale measurements. The extinction corrected
O magnitudes will be more accurate. 

The k-corrected radio loudness  parameter R$^{*}$, as shown in Equation 3, is the ratio of radio flux to optical flux. AGN are often described as radio-loud or radio-quiet depending on the value of this ratio. In terms of the R$^{*}$ parameter radio-loud objects have the value R$^{*} \geq 10$ ( $\log R^{*} \geq 1$) and radio-quiet objects have values of 
R$^{*} \leq 1$( $\log R^{*} \leq 0$). NLS1 galaxies have traditionally been found to be radio-quiet.

We found that only three of the objects have published radio data in other
useful wavelengths. We used data from the Green Bank 6 cm radio survey
to calculate the slopes for those three objects but we assumed a slope
of $\alpha_{radio} = -0.5$ in the form $\rm S_{\nu} \propto
\nu^{\alpha_{radio}}$ for the rest of the sample. This is the slope
that was used by \cite{hop01} in their study of the radio properties
of Seyfert 1 nuclei. It is possible that we are
underestimating the number of radio loud sources in our sample.

\section{Estimating the Black Hole Masses}
Recent papers have investigated the possibility that the unusual
spectral properties of NLS1 galaxies can be explained by assuming an
accretion rate close to Eddington and a small black hole mass. A small
black hole mass accreting efficiently could explain narrow permitted
line widths (by way of smaller Keplerian velocities) and lower radio
luminosities (assuming the black hole powers radio emission). We estimate
black hole masses of our sample using recent techniques to test
whether or not our data fits this paradigm.

To estimate the black hole masses we use the method of \cite{mclure}
which utilizes the FWHM(H$\beta$) and 5100 \AA $ $ luminosity as
surrogates for black hole mass. As mentioned previously, our spectra are not necessarily photometric, so they needed to be corrected before using this formula to estimate black hole mass. 
We used the POSS E and O magnitudes along with the IRAF package SYNPHOT to calibrate our spectra and find the appropriate correction factor. Briefly, armed with the transmission curve for POSS E and O plates, we send our spectra through the model O and E filters and calculate the the O and E magnitudes using SYNPHOT. We use the difference between these synthetic magnitudes and the measured, photometric magnitudes from the actual POSS plates to find the flux difference via $\Delta m \propto \log( \frac{f1}{f2})$. The correction is applied and the corrected flux used to determine $L_{5100 }$.

The formula we used is the following
from \cite{gru04}:

\begin{equation}
  \log{M_{BH}} = 5.17 + \log{R_{BLR}} + 2[\log{\rm FWHM(H\beta)} - 3]
\end{equation}

where

\begin{equation}
\log{R_{BLR}} = 1.52 + 0.70(\log \lambda L_{5100} - 37).
\end{equation}

\begin{figure}
\plotone{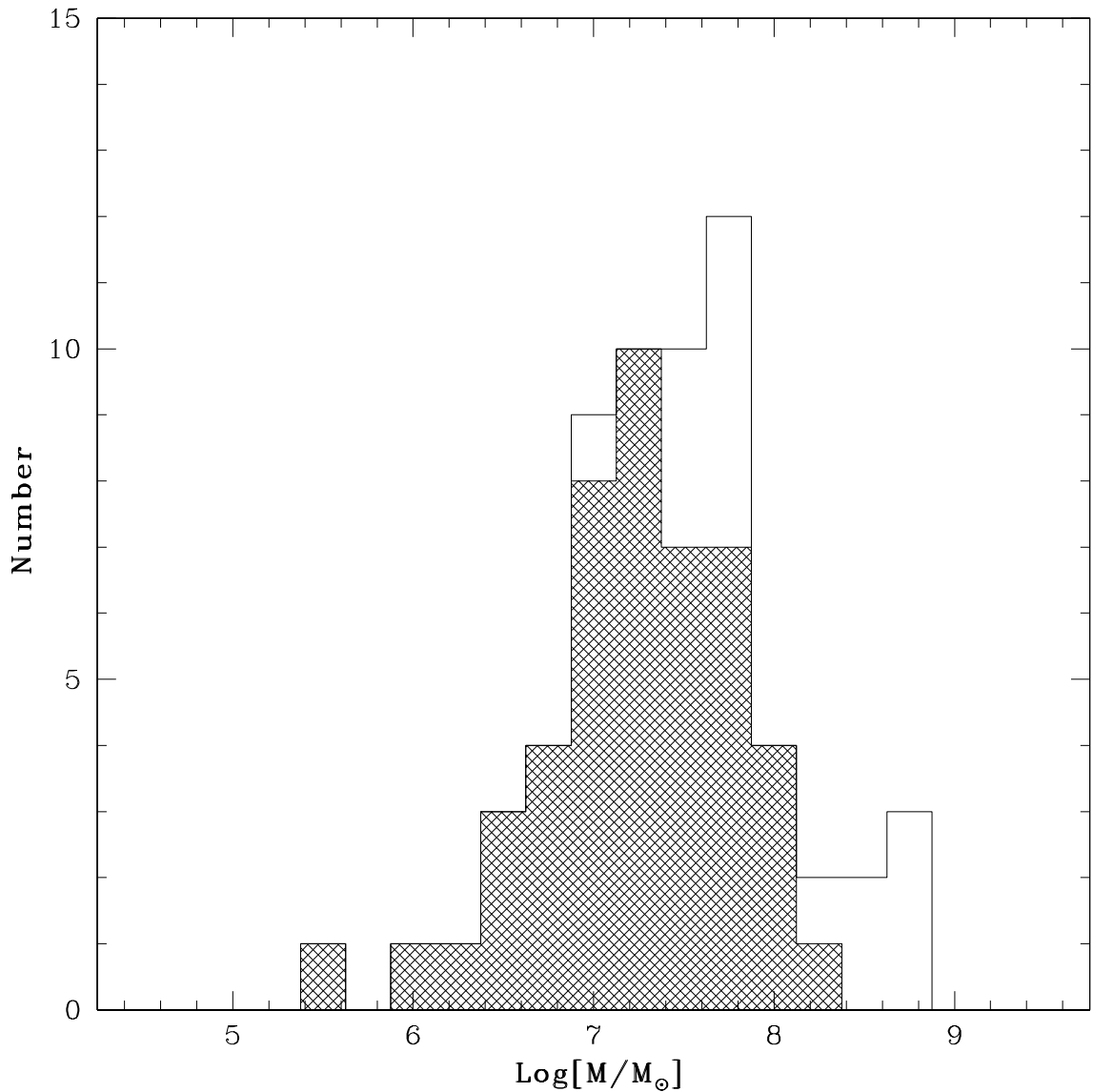}
\caption{Estimated black hole mass distribution for all 62 objects in
the sample using an empirical relation found by \cite{kas00}. The
masses are in units of solar masses. The shaded portion is the NLS1
galaxy distribution, the open region represents the BLS1 galaxy distribution.
\label{fig5}}
\end{figure}

Here $M_{BH}$ is the black hole mass in units of solar mass,
$M_{\sun}$, $R_{BLR}$ is the radius of the broad line region in
light-days, and $\lambda L_{5100}$ is the rest frame monochromatic
luminosity at 5100 \AA $ $ in Watts. Figure \ref{fig5} shows the
distribution of black hole masses for our sample, with the NLS1
galaxies being represented by the shaded parts of the histogram. These
masses are small for typical AGN.

\begin{figure}
\plotone{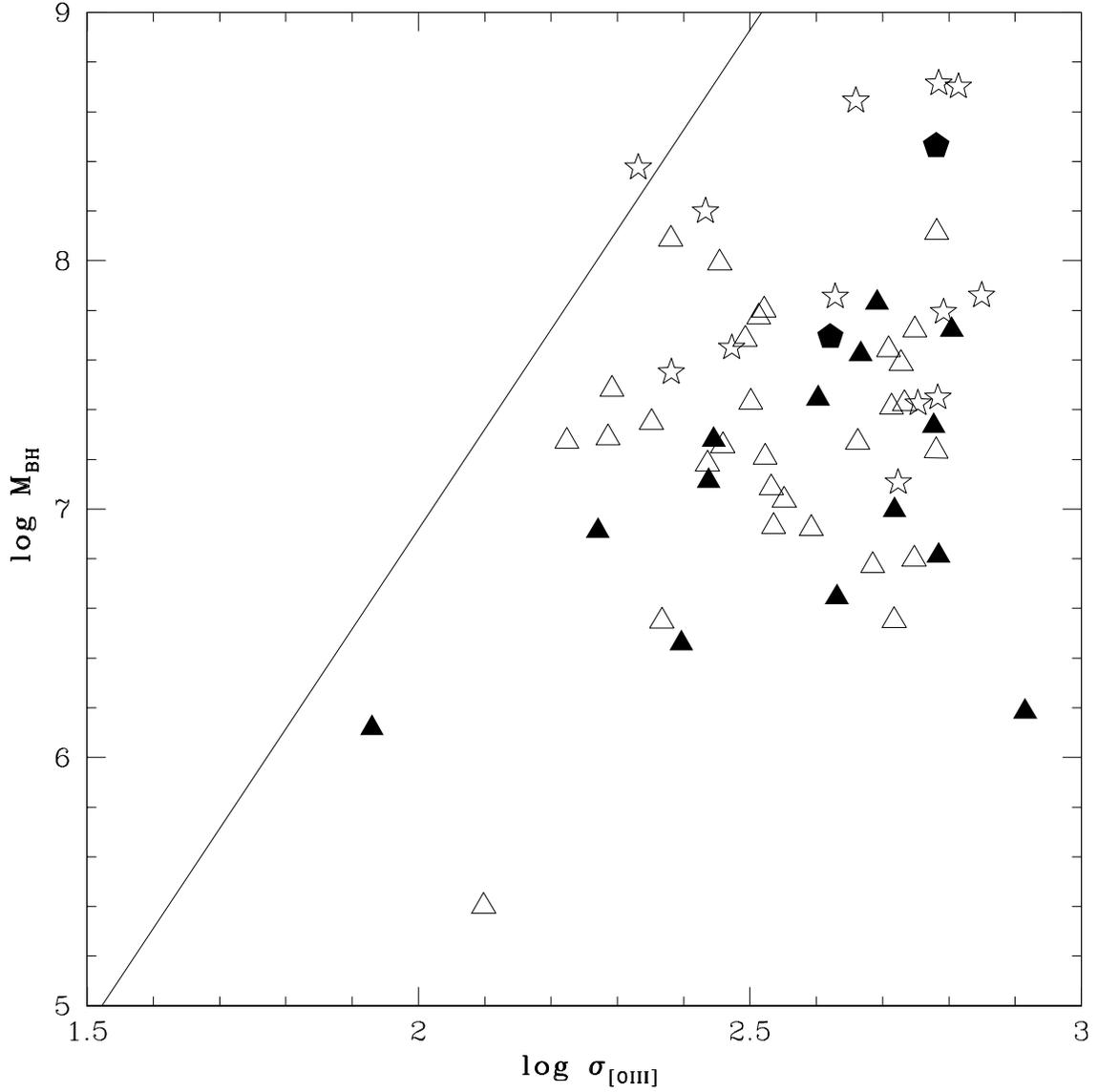}
\caption{$M_{BH}-\sigma$ relation for 59 of the 62 objects in our
sample. The open triangle are radio quiet NLS1s, filled trianges radio
loud NLS1s, the stars are radio quiet BLS1, and the two filled
pentagons represent radio loud BLS1s. The three missing objects have
their $\lambda 5007$ lines destroyed by atmospheric absorption, and
therefore were not included in this figure. The line is the relation
found by \cite{tre02} for normal galaxies.
\label{fig6}}
\end{figure}

Recent work by \cite{gru04} and \cite{mat01} suggests that NLS1
galaxies may be AGN in an early stage of AGN development. In this scenario, NLS1 galaxies have small
black hole masses that steadily grow by accretion in well formed bulges. To
illustrate this claim, the $M_{BH}-\sigma$
relation which is the tight correlation between central black
hole mass and bulge mass found by \cite{geb00}, \cite{fer00} among
others, is applied. The stellar velocity dispersion, which is used to measure the bulge mass, can be estimated using FWHM([OIII]) as was found in \cite{bor03} and references therein. Using the above formula for $M_{BH}$ and
FWHM([OIII]) as a surrogate for stellar velocity dispersion, the
$M_{BH}-\sigma$ relation for NLS1 galaxies was plotted. In their 2004 paper,
 Grupe \& Mathur applied this method to an X-ray selected sample of NLS1 galaxies
 and reported that the sample had both small black hole masses and that the galaxies fall
below the empirical line found by \cite{tre02} for normal
galaxies. They conclude that NLS1 galaxies do not follow the $M_{BH}-\sigma$
relation. As can be seen in Figure \ref{fig6}, although our data are
radio selected as opposed to X-ray selected, the radio selected NLS1
galaxies follow the same trend. Our NLS1 galaxies appear to have fairly small
black hole masses and do not closely follow the black hole mass-bulge
mass relation.

\section{Results and Correlations}
In this section we will calculate several parameters relevant to NLS1
study, test for correlations between sets of parameters, and then
compare with other published results. We run correlation tests with
several different cuts. Once with all objects in our sample included, once with
the BLS1 galaxies (FWHM(H$\beta) > 2000$ km s$^{-1}$) removed, and on each of the 
previous samples with 
radio loud and radio quiet points considered separately. We use the Spearman
rank correlation test because it is less sensitive to the occasional extreme data point.

\subsection{EW(FeII) vs. EW([OIII])}
\begin{figure} 
\plotone{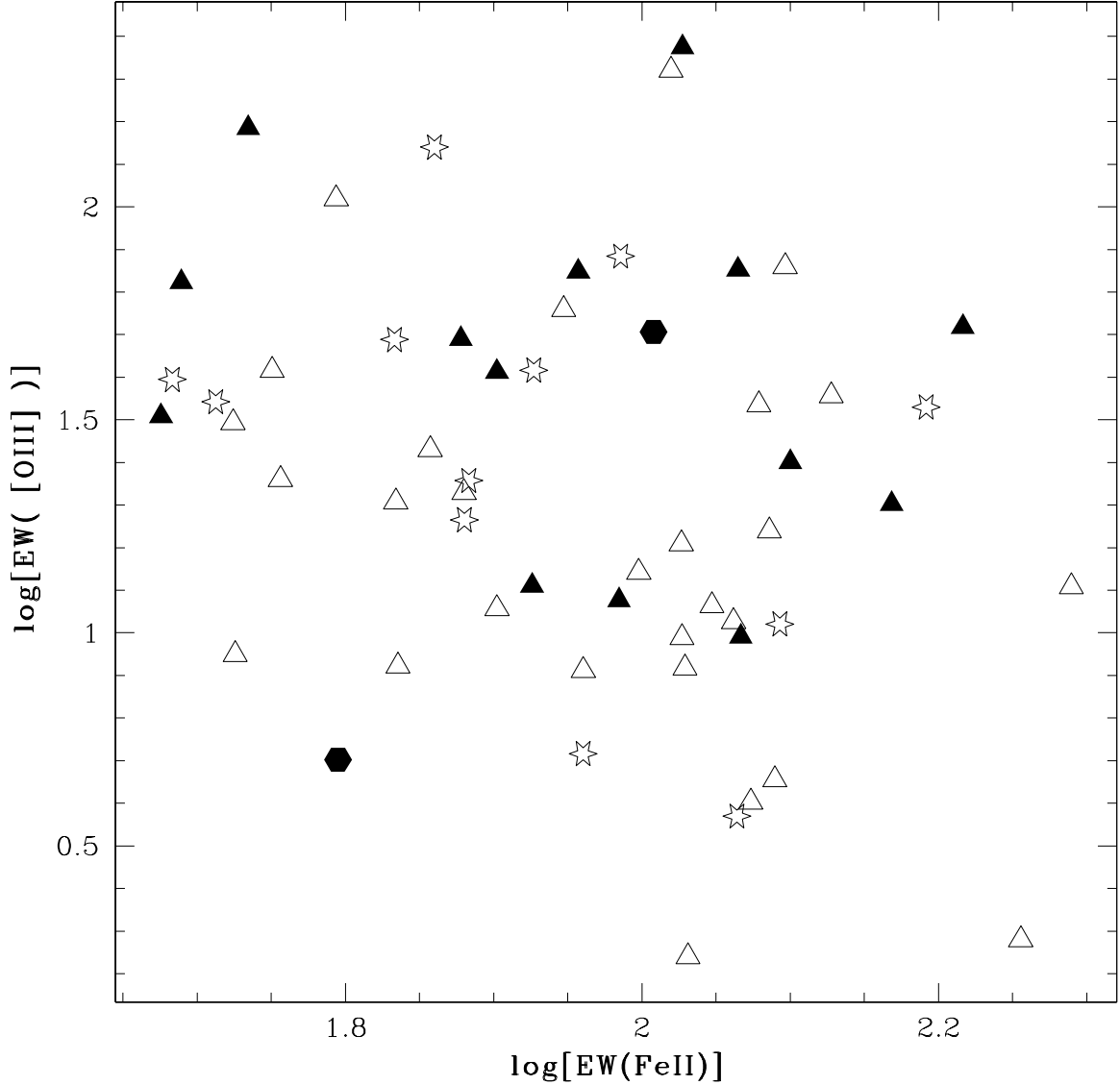}
\caption{A plot of the FeII and [OIII] equivalent widths. Open
triangles are radio quiet NLS1s, filled triangles are the radio loud
NLS1s, open stars are radio quiet BLS1s, and the filled pentagrams are
radio loud BLS1s. When all 56 objects that have these measurements
are considered, we find a relatively weak correlation,with $\rho_{Spearman} =
-0.250$ and a probability of no correlation of $0.064$. When only
NLS1s are considered the probability of no correlation increases to
$0.12$. A  correlation appears unlikely.
\label{fig7}}
\end{figure}

\cite{bor92} find that PC1 is driven by an anti-correlation between
the strengths of the iron complexes and the strengths of the
[OIII]$\lambda 5007$ line. We check in Figure \ref{fig7} for the same
correlation in our sample. For our sample we find a very weak borderline
anti-correlation ($\rho_{Spearman} = -0.250$, P $= 0.064$). These
correlations were performed on only 56 pairs of points as three of the
objects did not have measureable FeII and another three had [OIII]
lines eliminated by atmospheric absorption. When the test is performed
on only the NLS1 objects, the anti-correlation non-existent, with
$\rho_{Spearman} = -0.243$ and P$ = 0.119$. The trend found by Boroson
\& Green for low redshift quasars is not found in our sample of NLS1
galaxies. However, it is the case that, in general, this sample has
fairly weak [OIII] emission and strong FeII.

\subsection{EW([OIII]) vs. FWHM($H\beta$)}
\begin{figure} 
\plotone{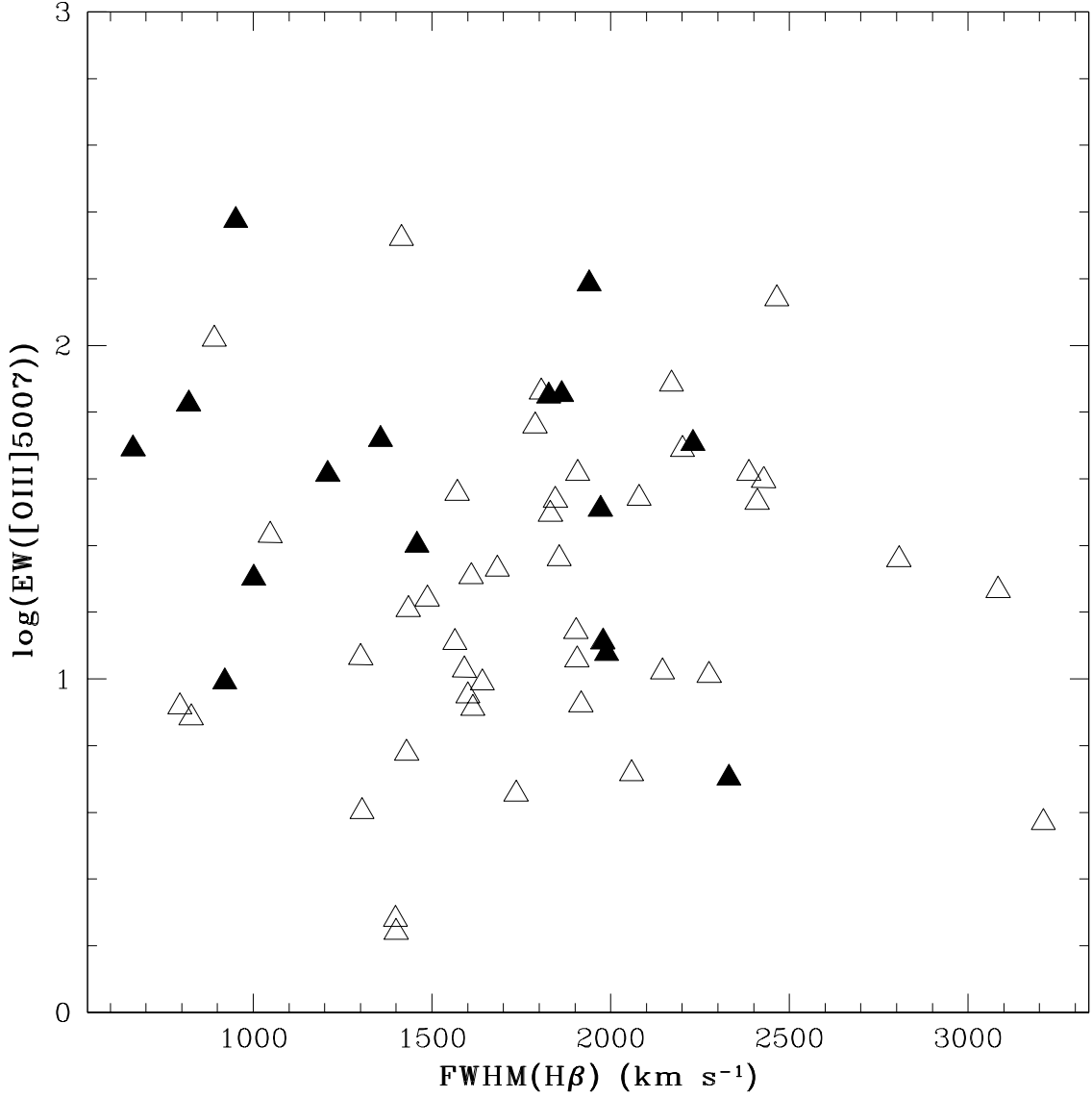}
\caption{The equivalent width of [OIII]$\lambda 5007$ measured in \AA $ $
plotted against the FWHM of H$\beta$ in km s$^{-1}$. The open
triangles represent NLS1 galaxies while the filled triangles represent
the BLS1 galaxies. There is no correlation between these two
parameters for the FBQS NLS1 galaxies. Spearman's test gives a
probability of chance correlation of P$ = 0.73$.
\label{fig8}}
\end{figure}

In Boroson \& Green's analysis they find that while PC1 is driven by
the anti-correlation between the strength of the FeII and [OIII]
lines, they also find that the equivalent width of [OIII] and
FWHM(H$\beta$) increase with PC1 and, therefore, are at least weakly
correlated. We test this possibility on the present sample in Figure
\ref{fig8}, where we plot the logarithm of EW([OIII]) against $\log
\rm FWHM(H\beta$). Our result does not confirm this statement as there appears to be no correlation between the two. Applying
the Spearman rank correlation test we get $\rho_{Spearman} = 0.045$
corresponding to a probability of chance correlation of $P = 0.73$
using a total of 59 data points. When we remove all the objects with
corrected FWHM(H$\beta$) $> 2000$ km s$^{-1}$ (leaving a total of 44
objects) and run the Spearman test again we get $\rho_{Spearman} =
0.040$ corresponding to a probability of $P = 0.795$, i.e., these
parameters are uncorrelated. The correlation between these two
parameters seen by Boroson \& Green is not present in our sample.

\subsection{EW(FeII) vs. [OIII]/H$\beta$ ratio}
\begin{figure} 
\plotone{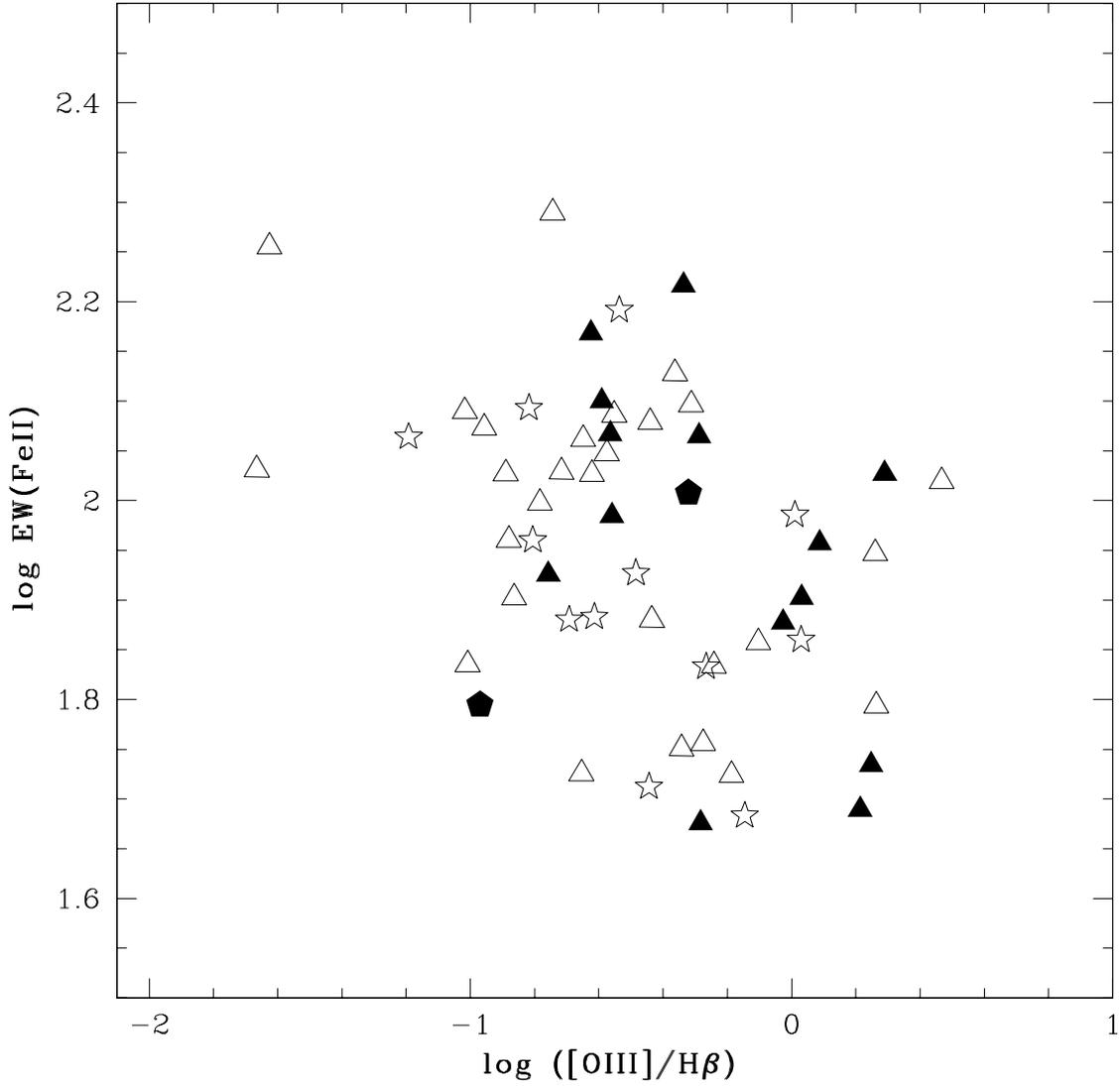}
\caption{FeII equivalent widths plotted against [OIII]/H$\beta$
ratio. EW(FeII) is measured in units of \AA. The open triangles are radio quiet
NLS1 galaxies, the filled triangles radio loud NLS1 galaxies, the
open stars are radio quiet BLS1 galaxies, and the filled pentagrams are
radio loud BLS1 galaxies. We integrated over two portions of each
individual spectrum to get the FeII equivalent width. The details are
in the text. The correlations are significant, with $\rho_{Spearman} =
-0.419$, P $= 0.007$ for the NLS1s and $\rho_{Spearman} = -0.39$, P$ =
0.004$ for the BLS1s.
\label{fig9}}
\end{figure}

In Figure \ref{fig9} we plot FeII equivalent width versus the
[OIII]/H$\beta$ ratio. The [OIII]/H$\beta$ always refers to the ratio
of the integrated rest-frame fluxes throughout this paper. Here there
is a significant anti-correlation with a spearman rank correlation
coefficient of $\rho_{Spearman} = -0.388$ and a probability of chance
correlation of $0.004$ for 56 of the objects in our sample. Of the six
objects excluded, three did not have [OIII] measurements as noted
earlier, and three others did not have a suitable FeII measurement. A
similar result holds for the 42 NLS1 galaxies, with $\rho_{Spearman} =
-0.419$ and P $= 0.007$. These are all noted in Table
\ref{tbl-2}. This is in agreement with the results of \cite{gru99},
which considers a sample of 76 soft X-ray sources with 36 NLS1
galaxies. This result was also found in other studies (e.g.,
\cite{bor92} and \cite{lao97}). Both the optically selected sample and
the X-ray selected sample show the same correlations as our radio
selected sample.

When the $\log R^* \ge 1$ and $\log R^* < 1$ samples are considered
separately, we find that the two sub-samples differ slightly. The
radio quiet ($\log R^* < 1$) sample is significantly correlated
($\rho_{Spearman} = -0.436$, P $=0.007$), in agreement with the result
for the larger sample. However, for the radio loud ($\log R^* \ge 1$)
sub-sample the correlation test fails, and no correlation is
present. As there were only 16 data points, and since Spearman is only
accurate for N $> 30$, we used a generalized Kendall's Tau method and
got $\tau_{Kendall} = -0.367$ with a corresponding probability of P $
= 0.32$. Regardless of the small number in the sub-sample, this result
should be accurate to within 5\%. There seems to be a significant
difference in the two sub-samples. In this case, the radio loud and
radio quiet NLS1 galaxies are behaving differently.

\subsection{ [OIII]/H$\beta$ ratio vs. optical slope}
\begin{figure} 
\plotone{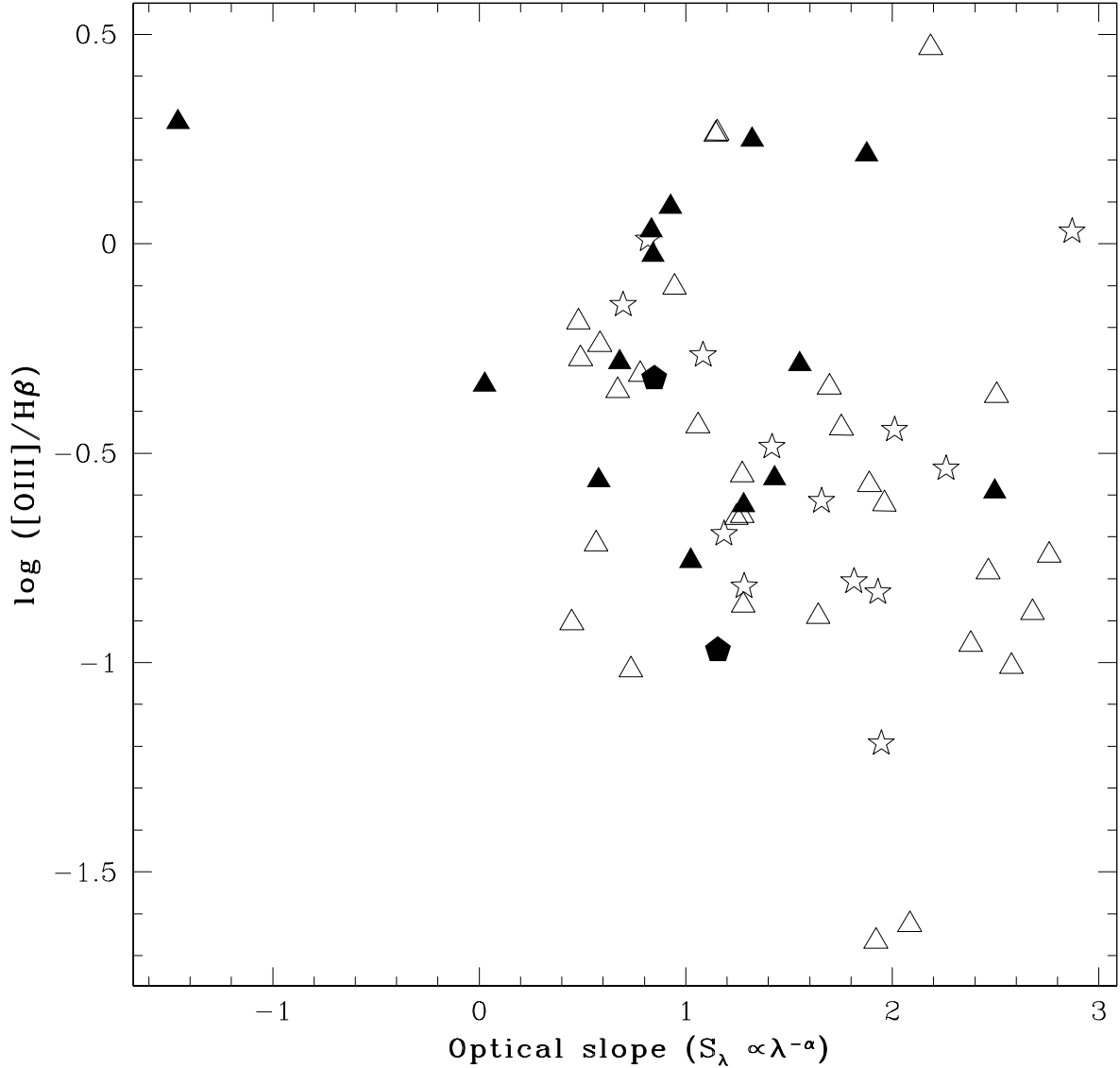}
\caption{[OIII]$\lambda 5007 / \rm H\beta$ vs. optical slope
$\alpha_{\rm opt}$. Open triangles represent radio quiet NLS1s, filled
triangles for radio loud NLS1s, open stars for radio quiet BLS1s, and
filled pentagrams for radio loud BLS1s. This is for a slope measured
between $3000$  \AA $ $ and $7000$  \AA. There is a statistically significant
anti-correlation between optical slope and [OIII]/H$\beta$ ratio with
$\rho_{\rm Spearman} = -0.345$ and probability of chance correlation
of $0.009$. There are only 59 data points as three of the objects have
no [OIII]$\lambda 5007$ measurement.
\label{fig10}}
\end{figure}

The plot of optical slope vs [OIII]/H$\beta$ ratio is shown in Figure
\ref{fig10}. We find a significant anti-correlation here, similar to
that found in \cite{gru99}, with a $\rho_{Spearman} = -0.345$ and a
chance probability of P $= 0.009$. The correlation is equally strong
when only NLS1 galaxies are considered. The [OIII]/H$\beta$ ratio
decreases as the optical continuum grows steeper and more blue. As
noted in Table \ref{tbl-2}, three of our objects have both of the
[OIII] lines rendered unmeasureable by atmospheric absorption.

 \subsection{EW(FeII) vs. $\log R^{*}$}
\begin{figure} 
\plotone{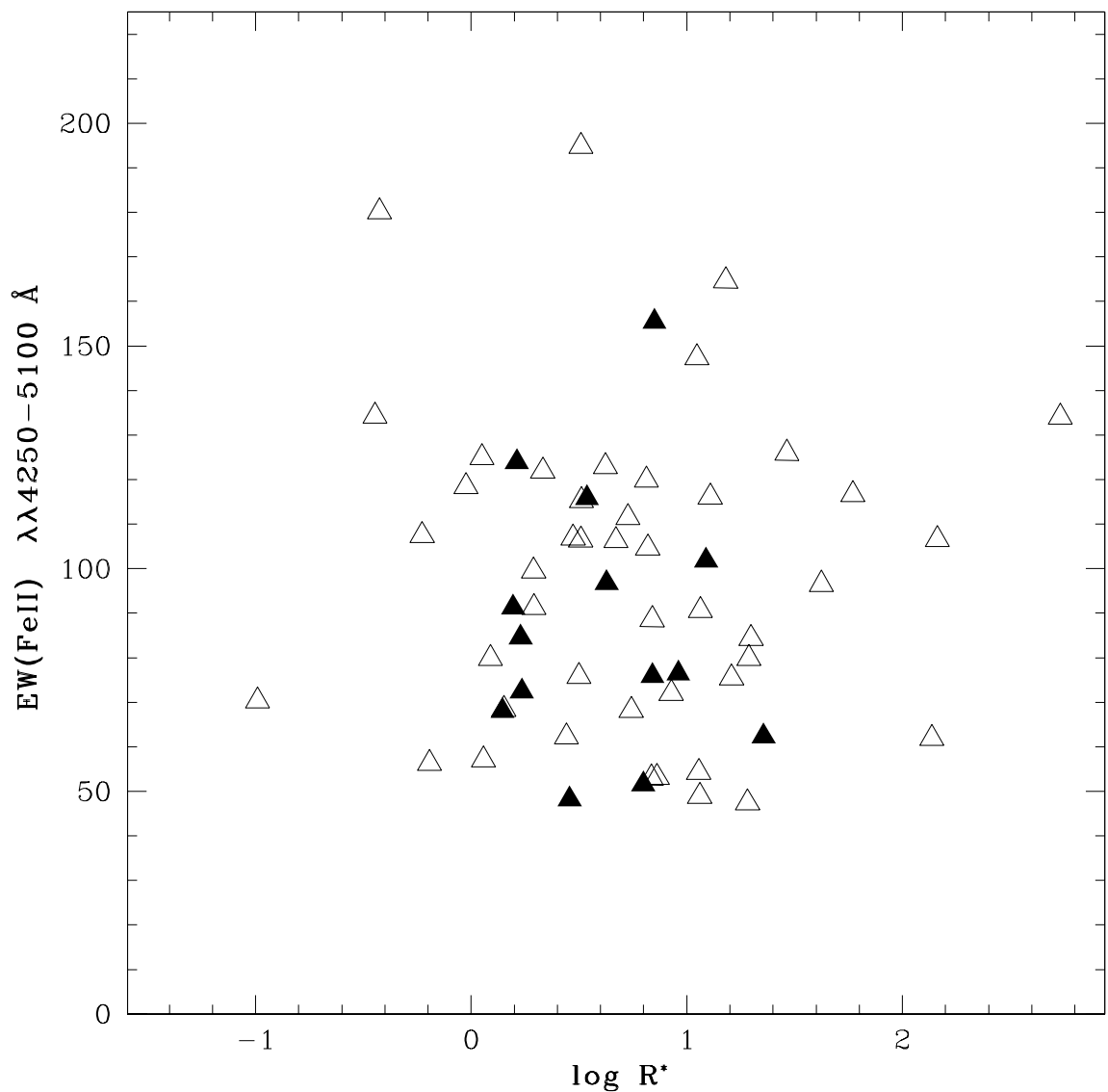}
\caption{Radio loudness paramenter $\log R^{*}$ and
EW(FeII). Here the open triangles are NLS1 galaxies while the filled
triangles are BLS1 galaxies. Boroson's PCA analysis for a sample of
162 optically selected AGN suggested the possibility that FeII
emission and radio loudness are linked. For the present collection of radio
selected NLS1 galaxies, this does not appear to be true.
\label{fig11}}
\end{figure}

It has been theorized that radio loudness can be correlated with FeII
strength, as FeII emission is thought to be directly related to
accretion rate. \cite{bor02} finds NLS1 galaxies at an extreme of PC1,
the interpretation of which is that FeII emission is strong when radio
loudness (and luminosity) is weak. This suggests the possibility that
FeII emission and radio loudness are linked in some way. We plot these
two quantities against one another in figure \ref{fig11} and
calculate the correlation strength. The result for the NLS1 galaxies is
$\rho_{Spearman} = -0.078$ and random probability of $0.61$. When the
BLS1 objects are added to the sample, the result is the same. We do not find a
correlation.

\subsection{FeII/[OIII] ratio vs. logR$^{*}$}
\begin{figure} 
\plotone{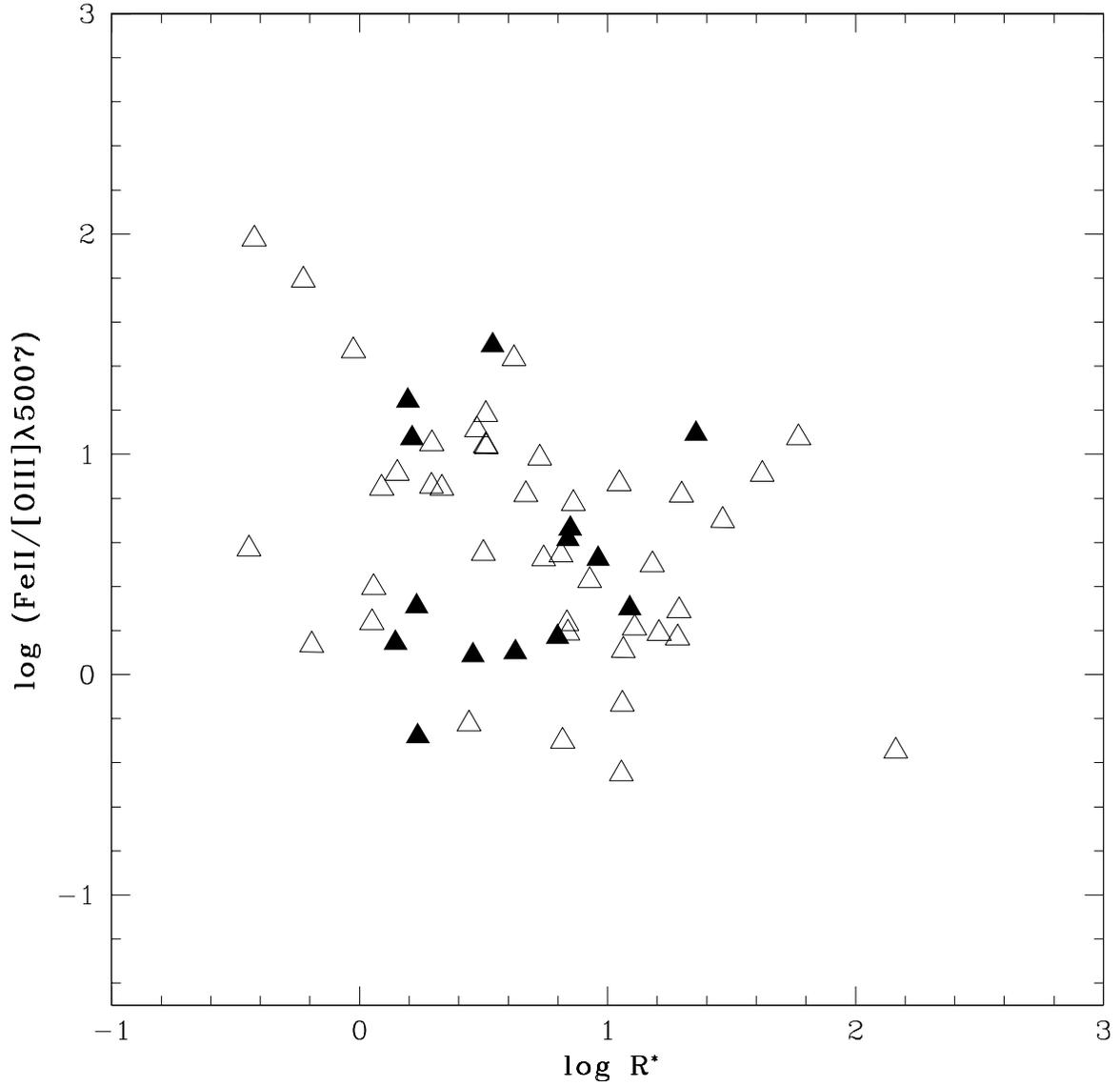}
\caption{FeII/[OIII] ratio vs. $\log{ \rm R^{*}}$. The open triangles
are NLS1 galaxies and the filled triangles are BLS1 galaxies. When
only the NLS1s are considered, the correlation is significant, with
$\rho_{\rm Spearman} = -0.387$ and $P = 0.013$. When all 56 points are
considered, the correlation is somewhat weaker with probability P $= 0.067$.
\label{fig12}}
\end{figure}

Figure \ref{fig12} shows the ratio of equivalent widths of the FeII
complexes and [OIII] 5007 emission vs. radio loudness. Boroson \&
Green reported for their optically selected sample of AGN that those
with stronger FeII emission and weaker [OIII] tended to be radio quiet
while those with weaker FeII and stronger [OIII] radio loud, though
not exclusively radio loud. Our radio selected sample of narrow and
broad line Seyferts suggests the same trend, with a weak
anti-correlation between FeII/[OIII] and radio loudness. Spearman's
rho for this plot is $\rho_{Spearman} = -0.254$ and probability of
random correlation is $0.059$. The NLS1 galaxies considered separately
have $\rho_{Spearman} = -0.387$ and P $= 0.013$, which is a noticeably
stronger correlation.

\begin{figure}
\plotone{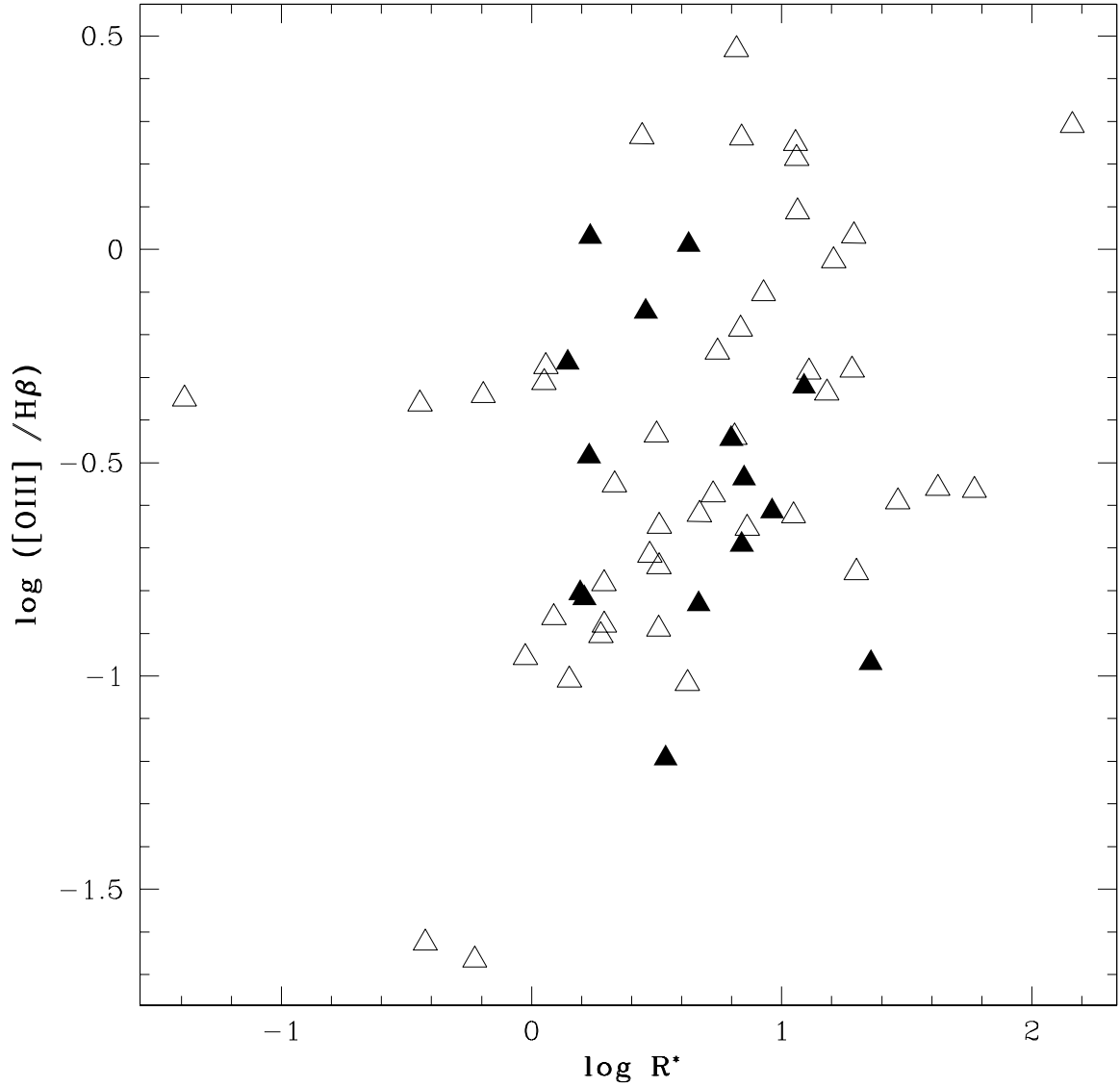}
\caption{[OIII]$\lambda 5007$/H$\beta$ ratio plotted against radio
loudness. The symbols are the same as in the previous figure, i.e, the filled triangles and NLS1 galaxies and open triangles are BLS1 galaxies. The Spearman rank
correlation test reveals a possible correlation, $\rho_{Spearman} =
0.302$ and chance probability of 2\%. When only NLS1 objects are
considered, the correlation is even stronger with $\rho_{\rm Spearman}
= 0.442$ and probability P $= 0.004$}
\label{fig13}
\end{figure}

In figure \ref{fig13} we have plotted logR$^{*}$ vs. [OIII]/H$\beta$
ratio. The Spearman correlation test on 44 NLS1 galaxies yields a
statistically significant correlation, with $\rho_{Spearman} = 0.442$ and probability of
no correlation of $0.004$. There is a trend toward higher logR$^{*}$
for greater [OIII]/H$\beta$ ratio. When the BLS1 galaxies are added,
the correlation is still fairly strong.

\subsection{FWHM(H$\beta$) vs $\log R^{*}$ and L$_{20 cm}$}
\begin{figure} 
\plotone{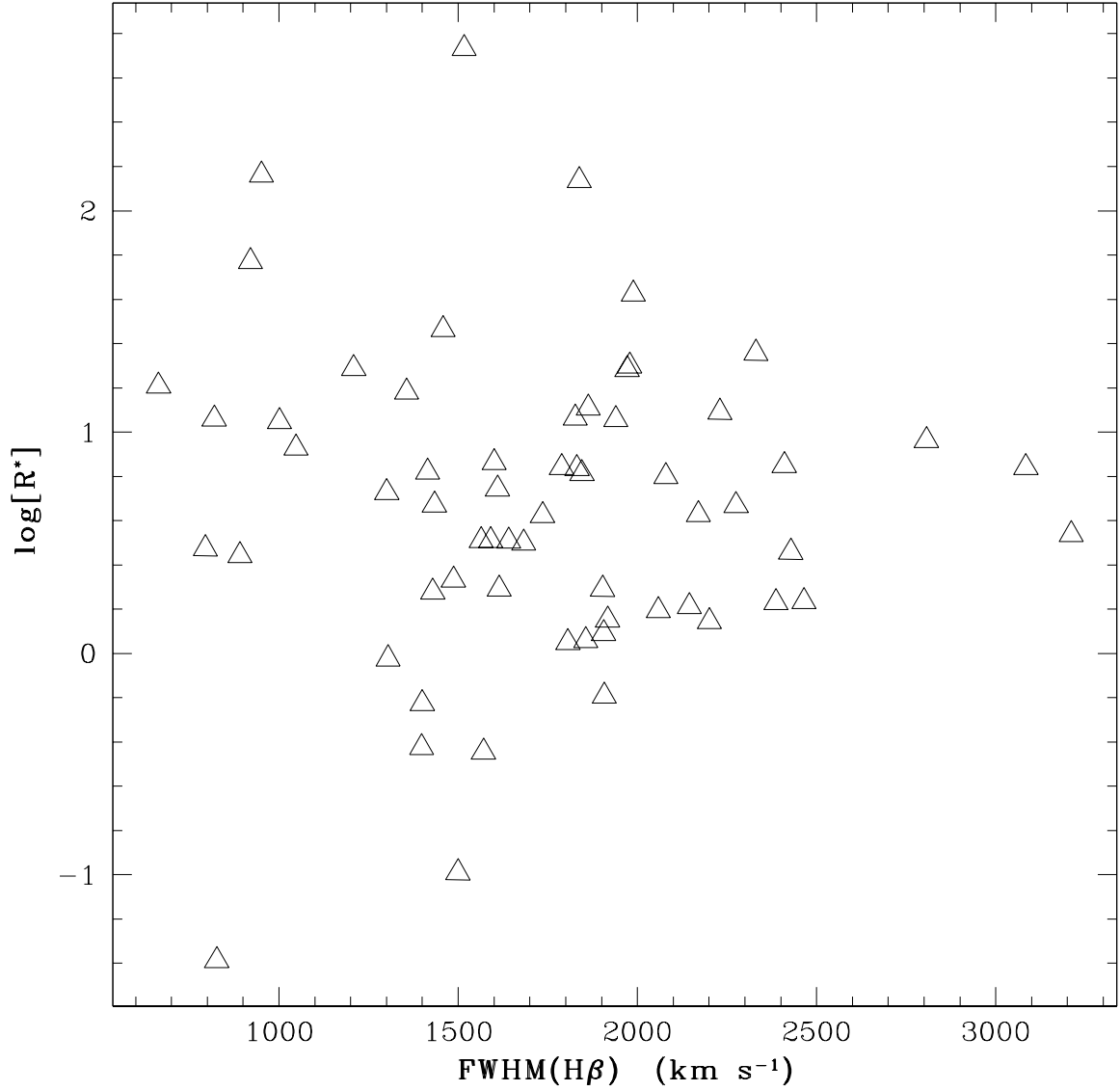}
\caption{FWHM(H$\beta$) (in km s$^{-1}$) vs. $\log R^{*}$ for the all 62 objects in this
sample. There is no correlation evident and Spearman rank correlation
tests confirm this with $\rho_{Spearman} = -0.046$ and P $= 0.72$  chance that
this correlation could happen with random data.
\label{fig14}}
\end{figure}

\begin{figure} 
\plotone{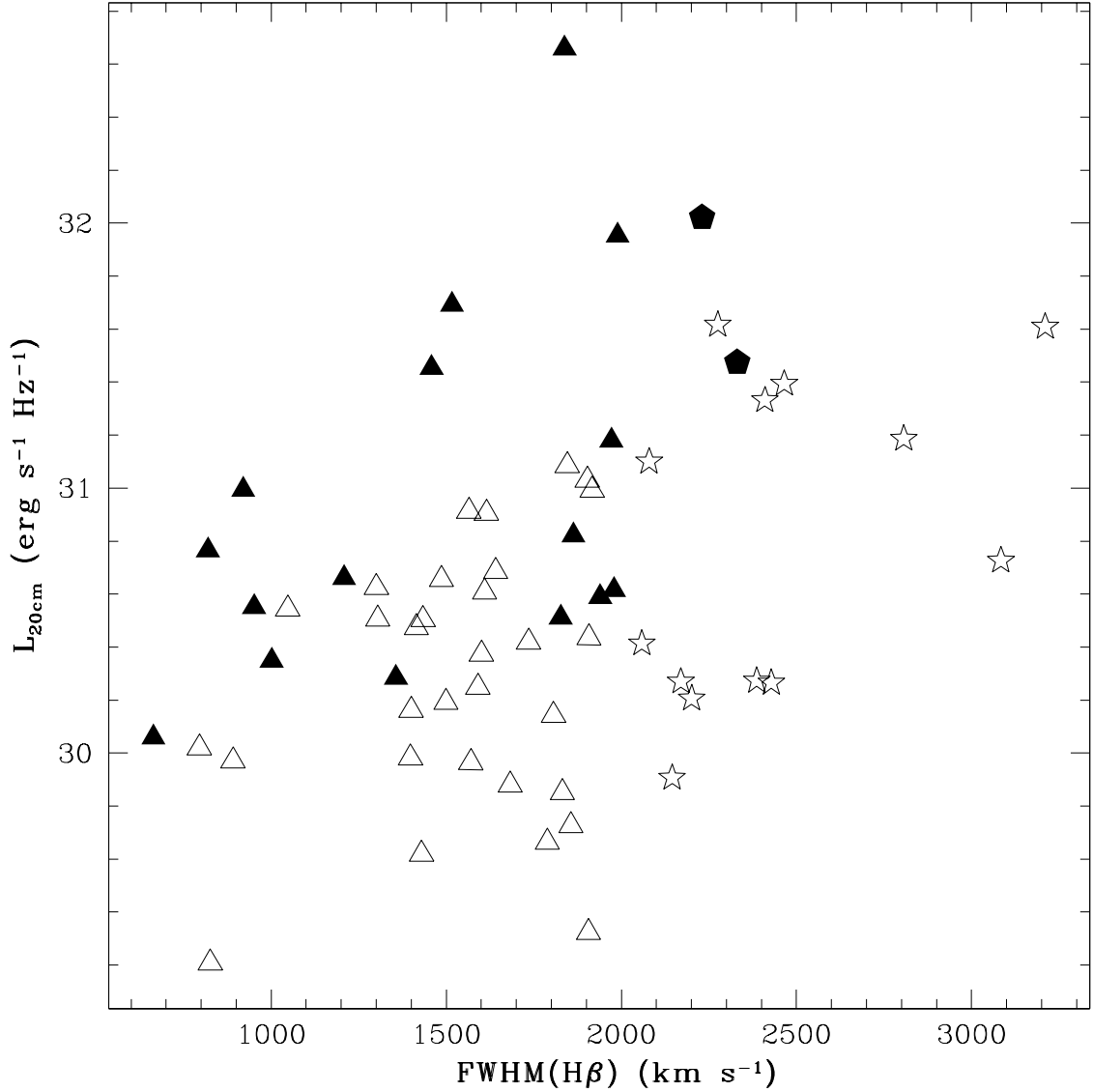}
\caption{Twenty centimeter radio luminosity plotted against
FWHM(H$\beta$). The filled symbols are the radio loud objects while
open symbols represent radio intermediate to radio quiet
objects. Triangles represent NLS1 objects. The other symbols are BLS1
objects. There is a statistically significant correlation ($\rho_{Spearman} = 0.345$, P
$= 0.007$) when all objects are considered. With only NLS1 objects
considered, the correlation is weaker, with $\rho_{\rm Spearman} =
0.246$ and probability P $= 0.095$. Vertical axis has units of erg
s$^{-1}$ Hz$^{-1}$ and the horizontal axis units of km s$^{-1}$.
\label{fig15}}
\end{figure}

In Figure \ref{fig14} we present a plot of radio loudness
vs. FWHM(H$\beta$)($\rho_{Spearman} = -0.046$, probability $= 0.72$),
which appears to be uncorrelated. We also test 20 cm radio luminosity
plotted against FWHM(H$\beta$). The Spearman rank correlation test was
run on all data points and the result was a significant correlation
with $\rho_{Spearman} = 0.331$ and probability of chance correlation
$= 0.01$. The results are shown in Figure \ref{fig15}. The radio loud
objects have filled in symbols and the others have open symbols. While
the radio loud objects are slightly more luminous in the radio than
the radio intermediate and radio quiet objects, they follow the same
trend; increasing FWHM(H$\beta$) follows increasing 20 cm radio power.

\section{Radio Properties}
\begin{figure} 
\plotone{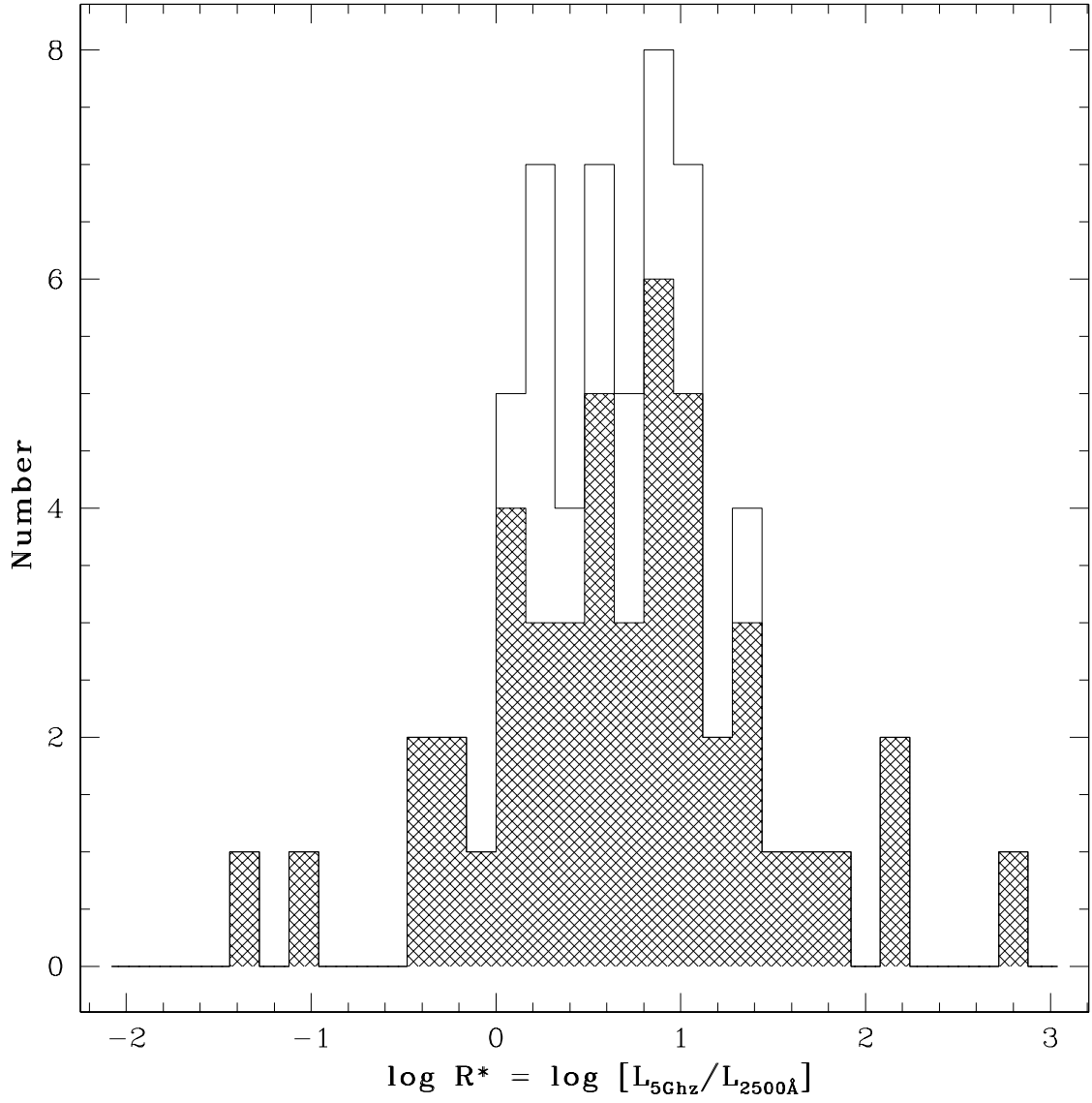}
\caption{Distribution of radio loudness parameter, $\log \rm R^{*}$,
for all 62 objects in our sample. The shaded portion represents the
NLS1 distribution and the open portion represents the BLS1 distribution. Our NLS1 distribution remarkably radio loud
compared with previous samples of NLS1 galaxies.
\label{fig16}}
\end{figure}

Probably the most interesting results from the analysis of our sample
are the radio properties. Until now, only 4 radio loud NLS1 galaxies
have been discovered and published \citep{rem86,gru00,osh01,zho03}. In
our sample we have 16, increasing the total number of known radio loud
NLS1 galaxies by a factor of $\sim 5$. The distribution of the
k-corrected radio loudness parameter, $\log R^{*}$, shown in Figure
\ref{fig16} demonstrates the singular nature of our sample. NLS1
galaxies are known to be radio quiet objects, most having little radio
emission or none at all. Note that the majority of the sources are
clustered between $0 < \log R^{*} < 1$. This distribution is typical
for FIRST sources, as the survey finds many objects in the radio
intermediate range. However, for a sample of NLS1 galaxies, the
distribution is quite striking in that it is very radio loud. Although
NLS1s are typically radio quiet, in this sample there are only a few
radio quiet objects. At this point there are only $\sim 350$ NLS1
galaxies published, four of which are radio loud. This sample adds
significantly to the number of known NLS1 galaxies and especially to
the number of radio loud NLS1 galaxies.

In figure \ref{fig17} we show several sample 20 centimeter radio
contours of our NLS1 galaxies. All the sources show little or no
structure, instead they are simply point sources. This agrees with the
argument put forth by \cite{goo89} that one possibility for the narrow
permitted line profiles in NLS1 galaxies is that they are seen pole
on, and that the gas motion is mainly confined to a disk. The lack of
structure is interesting, although many of these objects are only a
few mJy and more data are needed to confirm this result.

As stated previously, NLS1 galaxies are considered to be radio quiet and not
very radio luminous.  The only previous study of the radio properties
of NLS1 galaxies was presented in a 1995 paper by Ulvestad et al. (1995), who
looked at 15 NLS1 galaxies and detected 9 in the radio. They found radio powers
of $10^{27} - 10^{30}$ erg s$^{-1}$ Hz$^{-1}$ at 20 cm and all the objects
were radio quiet, i.e., $\log \rm R^{*} < 0$. The distribution of 20 cm
luminosities for our sample is shown in Figure \ref{fig18}. Our objects span
the range from $10^{29} - 10^{33}$ $ $ erg s$^{-1}$ Hz$^{-1}$, again, typical
for the FIRST survey but a factor of $10^{2} - 10^{3}$ times greater
than the Ulvestad study.

The \cite{bor92} statistical analysis of the spectral properties of 75
optically selected quasars from the Palomar-Green survey included a
small subsample of NLS1 galaxies. BG92 measured the optical
emission-line properties and a broad range of continuum properties,
from the X-ray through the radio, and performed a principal component
analysis (PCA) on the data. The PCA produced several eigenvectors, the
two most prominent being PC1 (related to Eddington ratio) and PC2
(related to accretion rate). When PC1 is plotted against PC2 their
figure seems to suggest that black hole mass correlates with radio
loudness. In figure \ref{fig19} we consider all 62 objects in our
sample and plot black hole mass vs. radio loudness and get
$\rho_{Spearman} = -0.18$ and probability of $P = 0.16$, so there
appears to be no correlation. Our results do not confirm a correlation
between radio loudness and black hole mass. When only the galaxies
with FWHM(H$\beta) < 2000$ km s$^{-1}$ are considered, the result is
$\rho_{Spearman} = -0.262$, P $= 0.076$. This is still a fairly weak
correlation.

NLS1 galaxies appear at the high Eddington ratio, low accretion rate extreme
of the PC1-PC2 diagram. According to this model, NLS1 galaxies have small
black hole masses and are radio quiet \citep{bor02}. We find that
while it is still probable that NLS1 galaxies are in general radio
quiet, they can have significant radio emission and can be radio
loud. We also find that in general, the central black hole mass for an
NLS1 galaxy is smaller than the typical AGN.

\subsection{Test for Radio Loud$/$Radio Quiet Dichotomy}

We wish to investigate what may be unusual about our radio loud NLS1
galaxies, to look for trends that could define what is different
between radio loud NLS1 and radio quiet NLS1 galaxies. As has been
shown by \cite{whi00}, there is no radio loud- radio quiet dichotomy
present in the FBQS. That property holds in the $\log R^{*}$
distribution of this sample of NLS1 galaxies, which, like the FBQS
study as a whole, has its peak in the radio intermediate range. We run
Student's t-test on the various parameters in our sample and look for
differences between radio loud and radio quiet samples. One of our
results is shown Figure \ref{fig20}. In this figure is shown the 
distribution of the [OIII]$\lambda 5007$$/$H$\beta$ ratio, which for
NLS1 galaxies is defined to be less than three. The two distribution
are $\log \rm R^{*} < 1$ (radio intermediate to radio quiet) in the
upper panel and $\log \rm R^{*} \geq 1$ (radio loud) in the lower
panel. The shaded portions represent the galaxies with FWHM(H$\beta) <
2000$ km s$^{-1}$ while the open portions are the FWHM(H$\beta) >=
2000$ km s$^{-1}$. The radio loud objects appear to be more heavily
weighted towards the high [OIII]-low H$\beta$ end of the ratio, while
the radio quiet objects appear to be dominated by larger H$\beta$
flux. The radio loud sample does appear to be somewhat different from
the radio quiet.

Student's t-test provides a more stringent test of whether or not two
sets of data differ significantly. When the test is performed on
[OIII]/H$\beta$ ratio data, separated into $\log \rm R^{*} \geq 1$ and
$\log \rm R^{*} < 1$, it suggests that for the NLS1 galaxies (44 total,
due to 3 missing [OIII] lines) the radio loud and radio quiet data
sets may very well be two different populations. The result is $t =
2.40$ and probability that the difference is due to chance is
$0.021$. A similar result is present in the radio luminosity L$_{20
cm}$(t $= 3.53$, P $= 0.01$). The result for radio luminosity strongly
suggests two different populations, but that is not unusual, as a
separation between radio loud and radio quiet should produce two
populations in radio measurements.

Performing Student's t-test in a similar manner on the FeII strength
data we get (t $= 0.07$, P $= 0.95$), meaning that the radio loud and
radio quiet populations are likely the same. A similar result holds
for both the EW(H$\beta$) measurements (t $=0.829$, P $= 0.41$) and
FWHM($H\beta$) (t $= 0.64$, P $= 0.50$). Again, there is no
significant difference between the radio loud and radio quiet
populations. These galaxies are probably not a new class of AGN, but
simply a group of regular NLS1 galaxies. However, they are
significantly more luminous in the radio than those found previously
and have greater [OIII] emission.

\section{Summary}
In this paper we presented the optical and radio properties of a
sample of radio selected Narrow-line Seyfert 1 galaxies. We find 16
new radio loud NLS1 galaxies to add to the three that were previously
discovered to increase the number of known radio loud NLS1 galaxies
significantly. We have run Spearman rank correlation tests between
several sets of parameters and find that, for the most part, our
results agree with previous studies of NLS1 galaxies. When compared
with X-ray selected and optically selected samples of NLS1 galaxies,
the properties of radio-selected NLS1 galaxies do not exhibit many obvious
 differences from NLS1 galaxy samples
selected in other manners. 
We also use Student's t-test to determine
whether radio loud NLS1 galaxies are statistically different from
their radio quiet counterparts and conclude that there are few
differences.

We estimate black hole masses with a relationship between
FWHM(H$\beta)$, optical luminosity, and M$_{BH}$. We also estimate
bulge velocity dispersion from FWHM([OIII]) and find that NLS1
galaxies do not obey the tight correlation between bulge mass and
black hole mass, confirming the result of \cite{gru04}.  Our sample
seems to agree with the canonical model of NLS1 galaxies, in that they
have small black hole masses and are accreting near the Eddington
rate. 

We have found a few
discrepancies between our sample and previous studies of NLS1 galaxies and other AGN.
We have found an anti-correlation between black hole mass and radio
loudness showing that as $M_{BH}$ increases, $\log \rm R^{*}$
decreases. In previous studies of AGN, it has been found that radio
loudness should increase as black hole mass increases. We test several
correlations from \cite{gru99} and \cite{bor02} and find that, while
many of the properties in our sample agree with theirs, there are a
few exceptions. We do not find an anti-correlation between the
strength of FeII and [OIII], nor do we find any connection between
FWHM(H$\beta$) and EW([OIII]). 

The most significant result found in this study is that we may now have to
consider that NLS1 galaxies can be strong radio emitters. Although the
galaxies in this study do not show the prodigious radio luminosities
of radio loud quasars, the luminosities we find are significantly
greater than the typical Seyfert galaxy. 

Our sample of radio selected NLS1 galaxies adds significantly to the
total of published NLS1 galaxies. Further investigation into this
growing class of radio loud NLS1 galaxies could help define what is
different physically between NLS1 and traditional Seyfert 1
galaxies. The authors are involved in a study of the X-ray properties
of the most radio luminous objects in this sample, which we hope will
reveal what is physically different about these NLS1
galaxies.

\acknowledgments This work was partly performed under the auspices of
the US Department of Energy, National Nuclear Security Administration
by the University of California, Lawrence Livermore National
Laboratory under contract No.  W-7405-Eng-48

\clearpage

\clearpage

\clearpage

\clearpage

\clearpage

\clearpage

\clearpage

\clearpage

\clearpage

\clearpage

\clearpage

\clearpage

\clearpage

\clearpage

\clearpage

\clearpage

\clearpage

\begin{figure} 
\plotone{f17.ps}
\caption{Twenty centimeter radio contours for the 12 radio loudest of
the FBQS NLS1 galaxies. Note that there is no obvious structure most
of the objects. They appear to be very compact, featurless radio
sources. The one exception is 1644+2619 which has a second hotspot
$\sim 11 \arcsec$ from the optical counterpart.
\label{fig17}}
\end{figure}

\clearpage

\begin{figure} 
\plotone{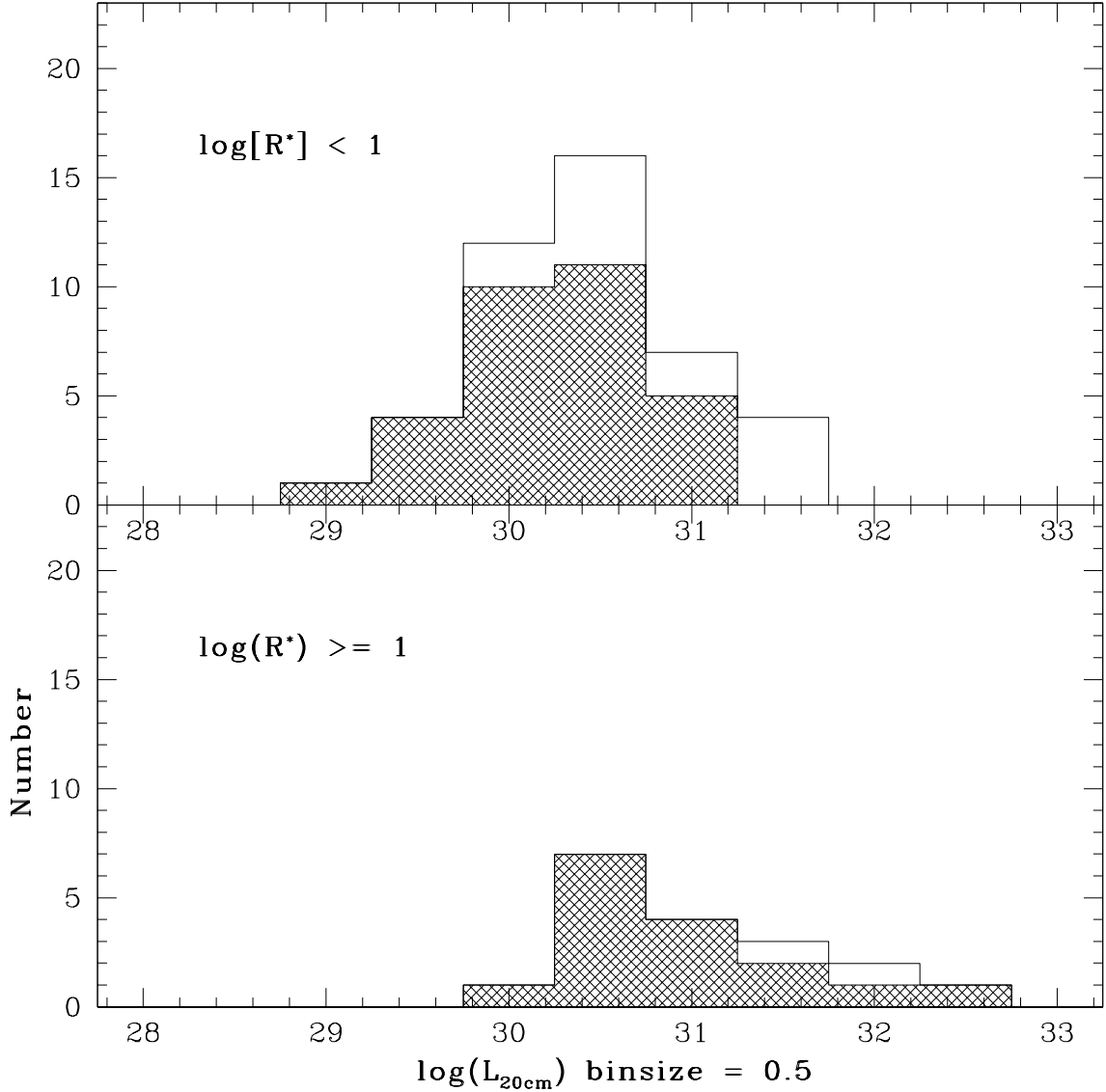}
\caption{Distribution of 20 cm radio luminosities for our sample. The shaded
portions represent the NLS1 galaxies and the open portion represents the BLS1 galaxies. The radio loud distribution is shown
in the bottom graph, while $\log \rm R^{*}< 1$ objects are on top.}
\label{fig18}
\end{figure}

\clearpage

\begin{figure} 
\plotone{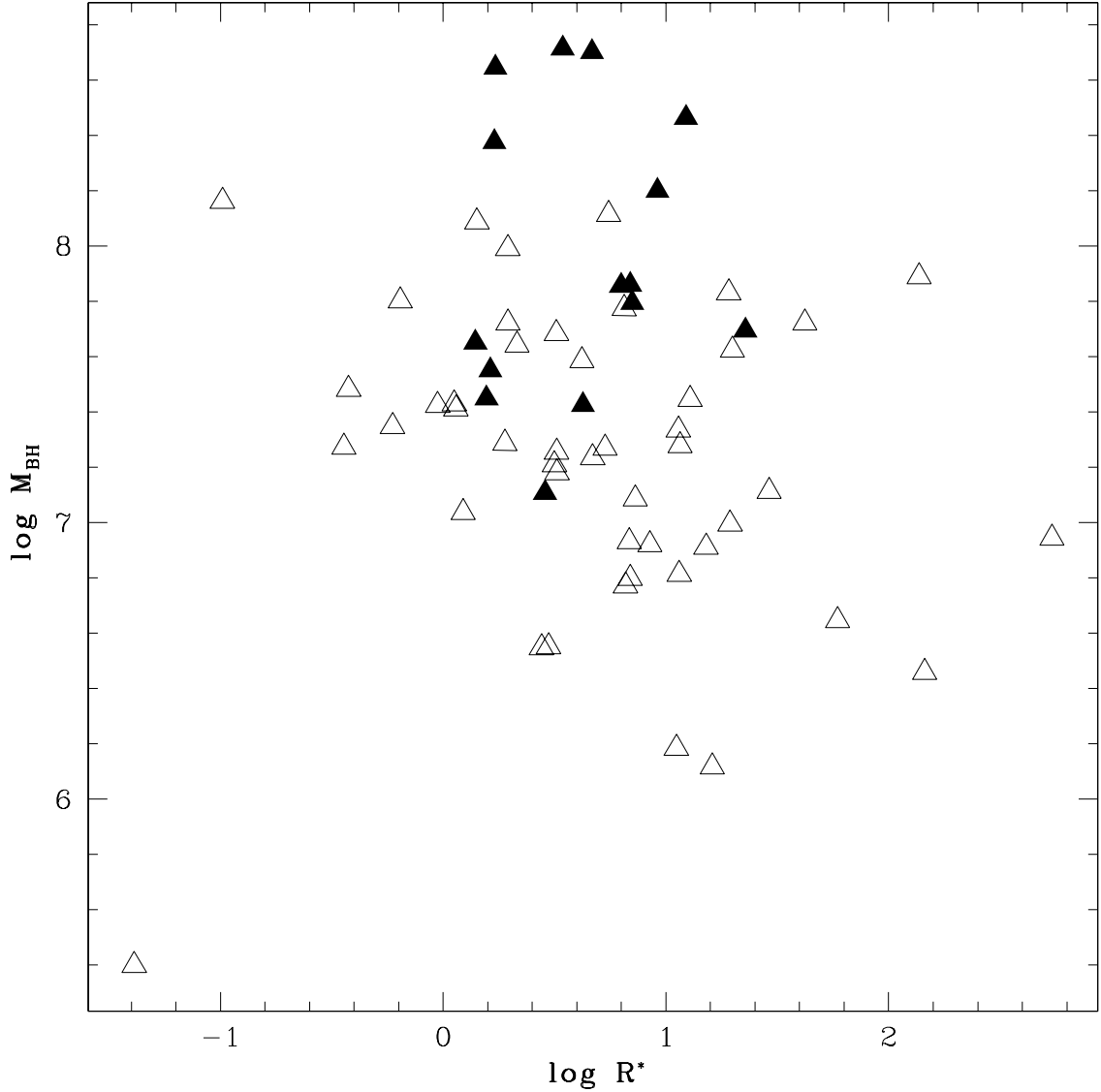}
\caption{Plot of black hole mass vs. $\log \rm R^{*}$. The open triangles are
NLS1 galaxies and the filled triangles represent the BLS1
galaxies. The masses are given in units of solar mass. The correlation
is not very compelling, with $\rho_{Spearman} = -0.18$ and the
probability of chance correlation being P $= 0.16$ for all 62 objects
in the sample. When considering just traditional NLS1 galaxies we get
a stronger correlation ($\rho_{Spearman} = -0.262$, P $= 0.076$) but
this is still not strong enough to get us tenure.
\label{fig19}}
\end{figure}

\clearpage

\begin{figure} 
\plotone{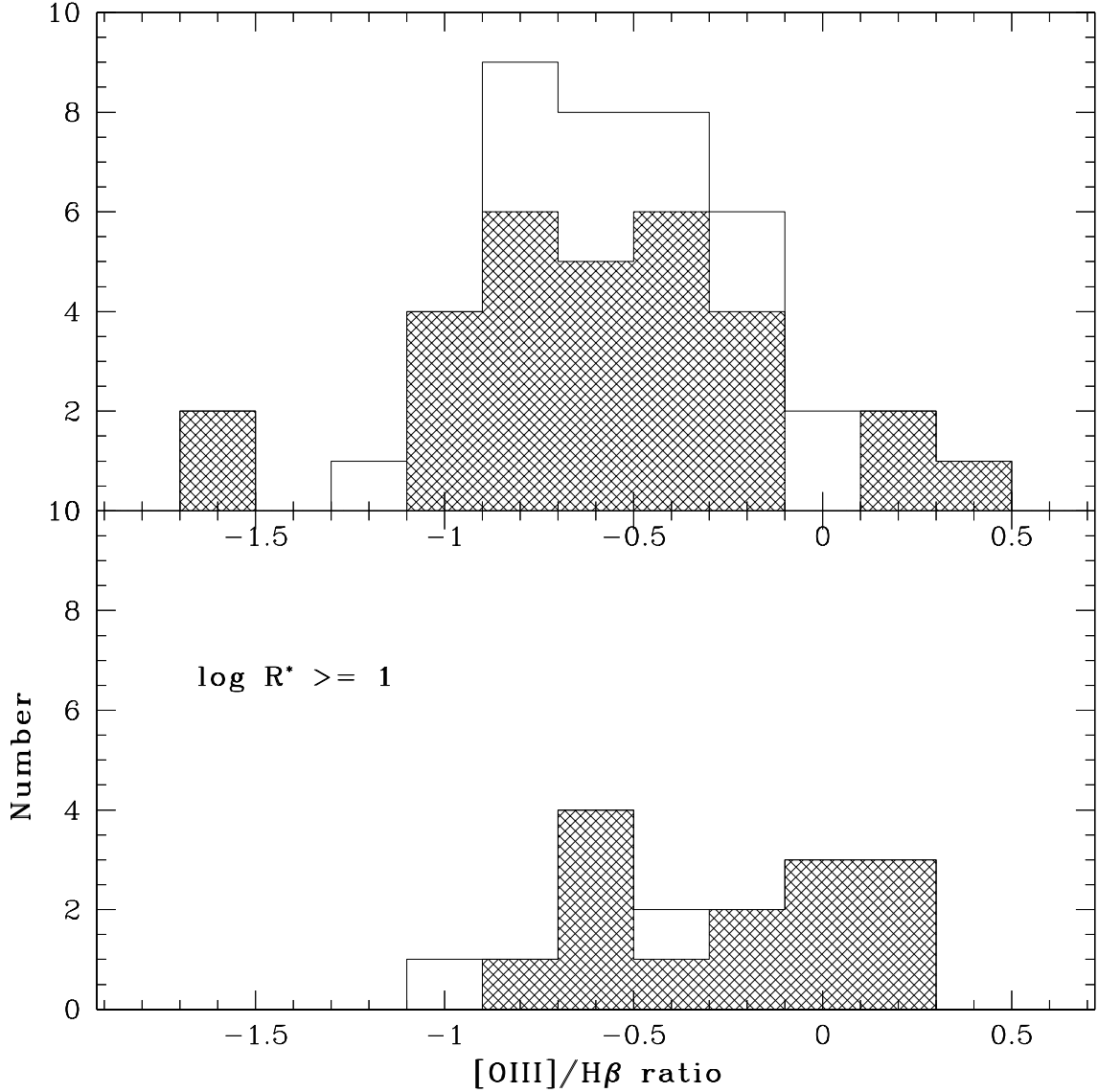}
\caption{The distributions of [OIII]/H$\beta$ ratio separated into a radio loud and a radio quiet population. We additionally isolate the NLS1s and BLS1s, with the shaded region
represents the NLS1 galaxies while the open region represents the BLS1
galaxies.The upper distribution is the radio quiet and the lower is
radio loud. There are a total of 59 objects in the two
distributions. Three objects with no [OIII] measurements are not
included.
\label{fig20}}
\end{figure}

\clearpage
\begin{deluxetable}{lcc}
\tabletypesize{\scriptsize}
\tablecaption{Spectrograph Characteristics \label{tbl-1}}
\tablewidth{0pt}
\tablehead{
\colhead{Telescope} & \colhead{Wavelength Range} & \colhead{Resolution}\\
\colhead{} & \colhead{\AA} &\colhead{\AA}}
\startdata
Lick 3 m & 3600$-$8150 & 6\\
KPNO 2.1 m & 3700$-$7400 & 4\\
APO 3.5 m & 3650$-$10000 & 10\\
MMT 6 $\times$ 1.8 m & 3600 $-$ 8500 & 8 \\
Keck II 10 m  & 3800$-$8800 & 8\\
ING 2.5 m & 4870$-$8280 & 3.33\\

\enddata

\end{deluxetable}

\clearpage

\begin{deluxetable}{lrrrrrrrrr}
\tabletypesize{\scriptsize}
\tablecaption{NLS1 data. \label{tbl-2}}
\tablewidth{0pt}
\tablehead{
\colhead{Source} & \colhead{redshift} & \colhead{$\log \rm R^{*}$} & \colhead{H$\beta$} & 
\colhead{[OIII]$\lambda$5007} &  
 \colhead{[OIII]/H$\beta$} & 
\colhead{EW(FeII)} & \colhead{EW(H$\beta$) } & 
\colhead{EW(OIII)} &  \colhead{slope ($\alpha$)}\\
\colhead{} & \colhead{} & \colhead{} & \colhead{FWHM} & \colhead{FWHM} & 
\colhead{} &  
\colhead{$\lambda\lambda$4434-4684 \AA} & \colhead{(\AA)} & \colhead{(\AA)} & \colhead{(S$_{\lambda} \propto \lambda^{-\alpha}$)}\\
\colhead{} & \colhead{} & \colhead{} & \colhead{(km s$^{-1}$)} & 
\colhead{(km s$^{-1}$)} & \colhead{} &  
\colhead{$\lambda\lambda$5147-5350 \AA} & \colhead{} & \colhead{} }
\startdata
0022$-$1039 & 0.414 &     0.81 & 1845 &  766 &    0.363 & 119.9 & 89.8 & 34.3 & 1.75 \\
0100$-$0200 & 0.227 &     1.77 &  920 & 1005 &    0.273 & 116.6 & 35.3 &  9.8 & 0.58 \\
0706+3901 & 0.086 &     1.21 &  664 &  200 &    0.941 & 75.5 & 50.8 & 49.0 & 0.84 \\
0713+3820 & 0.123 &     0.33 & 1487 & 1203 &    0.281 & 121.9 & 59.6 & 17.4 & 1.27 \\
0721+4329 & 0.157 &     0.21 & 2145 &  565 &    0.152 & 123.9 & 66.4 & 10.5 & 1.28 \\
0723+5054 & 0.203 &     1.29 & 1209 & 1228 &    1.074 & 79.8 & 37.3 & 41.0 & 0.83 \\
0729+3046 & 0.147 &     0.44 &  891 &  547 &    1.836 & 62.2 & 54.9 & 104.4 & 1.15 \\
0736+3926 & 0.118 &     0.05 & 1806 &  745 &    0.487 & 124.9 & 145.3 & 72.4 & 0.78 \\
0744+5149 & 0.460 &     1.62 & 1989 & 1498 &    0.276 & 96.5 & 41.5 & 11.9 & 1.43 \\
0747+4838 & 0.222 &     0.62 & 1735 & 1256 &    0.096 & 122.9 & 46.0 &  4.5 & 0.73 \\
0752+2617 & 0.082 &     0.09 & 1906 &  837 &    0.137 & 79.8 & 80.2 & 11.4 & 1.28 \\
0758+3920 & 0.095 &    $-$0.19 & 1908 &  780 &    0.455 & 56.3 & 86.5 & 41.4 & 1.69 \\
0804+3853 & 0.151 &     1.18 & 1356 &  438 &    0.461 & 164.6 & 113.3 & 52.2 & 0.03 \\
0810+2341 & 0.133 &     0.84 & 1831 &  807 &    0.650 & 53.0 & 47.3 & 31.2 & 0.48 \\
0818+3834 & 0.160 &     0.50 & 1683 &  784 &    0.367 & 75.8 & 56.3 & 21.3 & 1.06 \\
0833+5124 & 0.590 &     0.96 & 2806 &  637 &    0.243 & 76.4 & 89.2 & 22.8 & 1.66 \\
0909+3124 & 0.265 &     0.74 & 1610 & 1422 &    0.574 & 68.2 & 34.7 & 20.3 & 0.58 \\
0918+3024 & 0.538 &     0.85 & 2410 & 1455 &    0.291 & 155.5 & 109.1 & 33.9 & 2.26 \\
0937+3615 & 0.180 &     0.93 & 1048 &  919 &    0.788 & 72.0 & 33.2 & 26.9 & 0.95 \\
0946+3223 & 0.405 &     0.29 & 1615 & 1317 &    0.132 & 91.3 & 57.2 &  8.2 & 2.68 \\
1005+4332 & 0.179 &     0.19 & 2059 & 1427 &    0.156 & 91.2 & 31.5 &  5.2 & 1.81 \\
1010+3003 & 0.256 &    $-$0.02 & 1305 & 1272 &    0.111 & 118.4 & 33.8 &  4.0 & 2.38 \\
1038+4227 & 0.220 &     1.30 & 1979 & 1091 &    0.175 & 84.3 & 71.6 & 12.9 & 1.02 \\
1048+2222 & 0.329 &     0.73 & 1301 & 1080 &    0.266 & 111.5 & 41.3 & 11.6 & 1.89 \\
1102+2239 & 0.455 &     1.28 & 1972 & 1155 &    0.522 & 47.4 & 60.6 & 32.2 & 0.68 \\
1127+2654 & 0.379 &     0.29 & 1903 &  669 &    0.165 & 99.5 & 78.4 & 13.9 & 2.46 \\
1136+3432 & 0.193 &     0.23 & 918 &  504 &    0.328 & 84.5 & 121.2 & 41.4 & 1.42 \\
1140+4622 & 0.116 &     1.36 & 2331 &  982 &    0.107 & 62.3 & 45.4 &  5.0 & 1.16 \\
1145+2906 & 0.142 &     0.14 & 2201 &  698 &    0.542 & 68.1 & 87.2 & 48.8 & 1.08 \\
1151+3822 & 0.334 &     0.54 & 3210 & 1431 &    0.064 & 115.9 & 54.7 &  3.7 & 1.95 \\
1157+2613 & 0.324 &     0.84 & 3084 & 1662 &    0.203 & 75.9 & 87.8 & 18.4 & 1.19 \\
1159+2838 & 0.209 &     0.82 & 1415 & 1139 &    2.936 & 104.6 & 66.9 & 209.6 & 2.19 \\
1220+3853 & 0.377 &     0.15 & 1917 &  564 &    0.098 & 68.4 & 79.1 &  8.4 & 2.58 \\
1227+3214 & 0.137 &     2.16 &  951 &  585 &    1.951 & 106.5 & 126.8 & 236.9 & $-$1.46 \\
1256+3852 & 0.419 &     0.80 & 2079 &  999 &    0.360 & 51.6 & 91.3 & 34.9 & 2.01 \\
1313+3753 & 0.655 &     0.68 & 2275 & 1533 &    0.155 & \nodata & 62.7 & 10.1 & 1.94 \\
1333+4141 & 0.225 &     1.06 & 1940 & 1407 &    1.770 & 54.2 & 83.2 & 153.2 & 1.32 \\
1346+3121 & 0.246 &     0.86 & 1600 &  800 &    0.222 & 53.2 & 38.6 &  8.9 & 1.24 \\
1358+2658 & 0.331 &     1.11 & 1863 &  941 &    0.515 & 116.1 & 132.3 & 71.3 & 1.55 \\
1405+2555 & 0.165 &    $-$0.42 & 1398 &  460 &    0.024 & 180.1 & 75.6 &  1.9 & 2.08 \\
1405+2657 & 0.713 &     1.09 & 2230 & 1419 &    0.477 & 101.8 & 103.9 & 50.9 & 0.85 \\
1408+2409 & 0.131 &     0.51 & 1590 &  641 &    0.225 & 115.3 & 45.6 & 10.6 & 1.27 \\
1421+2824 & 0.540 &     2.14 & 1838 & \nodata & \nodata & 61.9 & 52.2 & \nodata & 2.18 \\
1431+3416 & 0.715 &     0.23 & 2465 & 1073 &    1.071 & 72.4 & 118.4 & 138.1 & 2.87 \\
1442+2623 & 0.108 &     0.47 &  795 & 1226 &    0.192 & 106.9 & 42.3 &  8.3 & 0.57 \\
1448+3559 & 0.114 &     0.06 & 1856 & 1216 &    0.531 & 57.0 & 42.5 & 22.9 & 0.49 \\
1517+2239 & 0.109 &     0.84 & 1789 & 1314 &    1.823 & 88.5 & 30.4 & 57.4 & 1.15 \\
1517+2856 & 0.209 &     0.63 & 2170 & 1332 &    1.024 & 96.8 & 73.1 & 76.7 & 0.82 \\
1519+2838 & 0.270 &     0.51 & 1641 &  731 &    0.129 & 106.4 & 72.0 &  9.7 & 1.64 \\
1610+3303 & 0.098 &     0.46 & 2428 & 1243 &    0.715 & 48.2 & 53.9 & 39.3 & 0.70 \\
1612+4219 & 0.233 &     1.06 &  819 & 1430 &    1.635 & 48.9 & 38.6 & 66.7 & 1.88 \\
1629+4007 & 0.272 &     1.46 & 1458 &  643 &    0.256 & 125.9 & 91.2 & 25.2 & 2.50 \\
1644+2619 & 0.145 &     2.73 & 1517 &  \nodata &   \nodata & 134.0 & 67.9 & \nodata & $-$0.11 \\
1702+3247 & 0.164 &    $-$0.23 & 1400 &  528 &    0.022 & 107.4 & 76.0 &  1.7 & 1.92 \\
1709+2348 & 0.254 &     1.06 & 1827 &  655 &    1.224 & 90.6 & 56.0 & 70.4 & 0.93 \\
1713+3523 & 0.085 &     1.05 & 1002 & 1933 &    0.237 & 147.4 & 81.4 & 20.1 & 1.28 \\
1716+3112 & 0.110 &    $-$0.45 & 1571 &  393 &    0.434 & 134.2 & 77.2 & 36.0 & 2.50 \\
1718+3042 & 0.281 &     0.67 & 1434 & 1418 &    0.239 & 106.3 & 63.8 & 16.2 & 1.96 \\
2155$-$0922 & 0.192 &    $-$0.99 & 1500 &  \nodata &   \nodata & 70.3 & 31.1 & \nodata & 2.03 \\
2159+0113 & 0.101 &     0.28 & 1429 &  454 &    0.124 & \nodata & 47.4 &  6.0 & 0.45 \\
2327$-$1023 & 0.065 &    $-$1.39 &  652 &  538 &    0.390 & \nodata & 7.66 & 16.85 & 0.67 \\
2338$-$0900 & 0.374 &     0.51 & 1564 &  676 &    0.180 & 194.8 & 65.7 & 12.8 & 2.76 \\
\enddata

\tablecomments{Objects 1421+2824, 1644+2619, and 2155-0922 had [OIII] emission lines destroyed by atmospheric lines.}

\end{deluxetable}

\end{document}